\documentclass[aps,natphysics,superscriptaddress,two column,balance,preprintnumbers,pagewise,notitlepage,bibnotes]{revtex4-2}
\usepackage{latexsym}
\usepackage{subcaption}
\usepackage{balance}
\usepackage{caption}
\usepackage[margin=0.75in]{geometry}
\usepackage[dvips]{color}
\usepackage[pdftex]{graphicx}
\usepackage{amsmath}
\usepackage{amssymb}
\usepackage{esdiff}
\usepackage{ragged2e}
\usepackage{lineno}
\usepackage{gensymb}
\usepackage{enumerate}
\usepackage{hyperref}
\usepackage{mathrsfs}
\usepackage[usenames,dvipsnames]{xcolor}
\usepackage{float}
\begin{document}
 
\title{\textbf{Spin-orbit coupling-enhanced valley ordering of malleable bands in twisted bilayer graphene on WSe$_2$}}
\author{Saisab Bhowmik}
\email{saisabb@iisc.ac.in}
\affiliation{\textit{Department of Instrumentation and Applied Physics, Indian Institute of Science, Bangalore, 560012, India}}
\author{Bhaskar Ghawri}
\affiliation{\textit{Department of Physics, Indian Institute of Science, Bangalore, 560012, India}}
\author{Youngju Park}
\affiliation{\textit{Department of Physics, University of Seoul, Seoul 02504, Korea}}
\author{Dongkyu Lee}
\affiliation{\textit{Department of Physics, University of Seoul, Seoul 02504, Korea}}
\affiliation{\textit{Department of Smart Cities, University of Seoul, Seoul 02504, Korea}}
\author{Suvronil Datta}
\affiliation{\textit{Department of Instrumentation and Applied Physics, Indian Institute of Science, Bangalore, 560012, India}}
\author{Radhika Soni}
\affiliation{\textit{Department of Instrumentation and Applied Physics, Indian Institute of Science, Bangalore, 560012, India}}
\author{K. Watanabe}
\affiliation{\textit{Research Center for Functional Materials, National Institute for Materials Science, Namiki 1-1, Tsukuba, Ibaraki 305-0044, Japan}}
\author{ T. Taniguchi}
\affiliation{\textit{International Center for Materials Nanoarchitectonics, National Institute for Materials Science, Namiki 1-1, Tsukuba, Ibaraki 305-0044, Japan}}
\author{Arindam Ghosh}
\affiliation{\textit{Department of Physics, Indian Institute of Science, Bangalore, 560012, India}}
\affiliation{\textit{Centre for Nano Science and Engineering, Indian Institute of Science, Bangalore 560 012, India}}
\author{Jeil Jung}
\email{jeiljung@uos.ac.kr}
\affiliation{\textit{Department of Physics, University of Seoul, Seoul 02504, Korea}}
\affiliation{\textit{Department of Smart Cities, University of Seoul, Seoul 02504, Korea}}
\author{U. Chandni}
\email{chandniu@iisc.ac.in}
\affiliation{\textit{Department of Instrumentation and Applied Physics, Indian Institute of Science, Bangalore, 560012, India}}



\begin{abstract}

New phases of matter can be stabilized by a combination of diverging electronic density of states, strong interactions, and spin-orbit coupling.~Recent experiments in magic-angle twisted bilayer graphene (TBG) have uncovered a wealth of novel phases as a result of interaction-driven spin-valley flavour polarization \cite{cao2018correlated, cao2018unconventional, lu2019superconductors, serlin2020intrinsic, sharpe2019emergent, bhowmik2022broken, das2021symmetry, wu2021chern}.~In this work, we explore correlated phases appearing due to the combined effect of spin-orbit coupling-enhanced valley polarization and large density of states below half filling ($\nu \lesssim 2$) of the moir\'{e} band in a TBG coupled to tungsten diselenide.~We observe anomalous Hall effect, accompanied by a series of Lifshitz transitions, that are highly tunable with carrier density and magnetic field.~Strikingly, the magnetization shows an abrupt sign change in the vicinity of half-filling, confirming its orbital nature.~The coercive fields reported are about an order of magnitude higher than previous studies in graphene-based moir\'{e} systems~\cite{serlin2020intrinsic, sharpe2019emergent, kuiri2022, chen2022tunable, polshyn2020electrical, lin_spin-orbit, chen2021electrically}, presumably aided by a Stoner instability favoured by the van Hove singularities in the malleable bands.~While the Hall resistance is not quantized at zero magnetic fields, indicative of a ground state with partial valley polarization, perfect quantization and complete valley polarization are observed at finite fields. Our findings illustrate that singularities in the flat bands in the presence of spin-orbit coupling can stabilize ordered phases even at non-integer moir\'{e} band fillings.

\end{abstract}

\maketitle

\begin{figure*}[bth]
\includegraphics[width=1.0\textwidth]{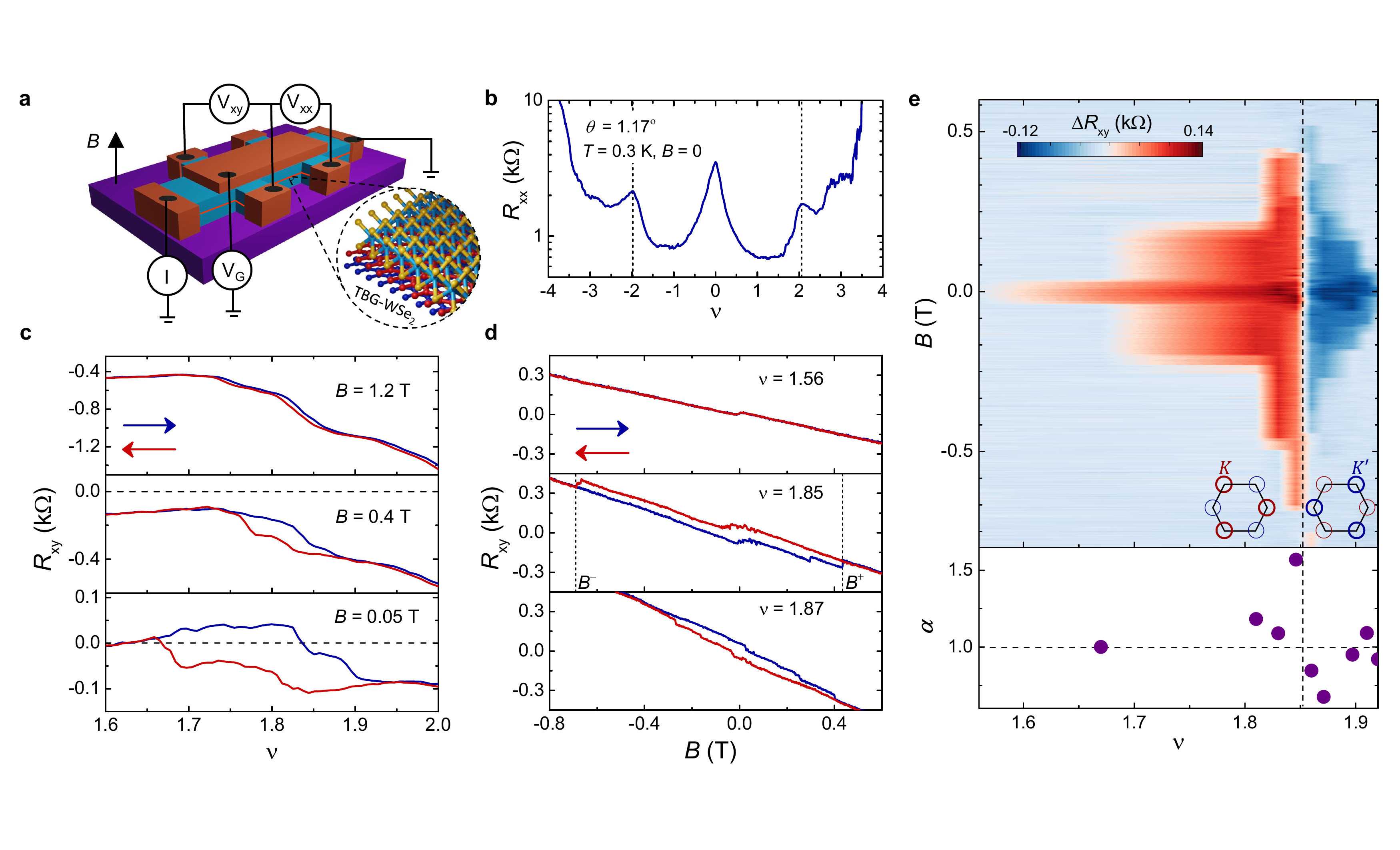}
\captionsetup{justification=raggedright,singlelinecheck=false}
\justify{\textbf{Fig.~1.~Ferromagnetism and valley polarization at $\boldsymbol{\nu<}~2$}.~\textbf{a.}~Schematic of hBN-encapsulated TBG-WSe$_2$ heterostructure on SiO$_2$/Si substrate.~\textbf{b.}~Four-probe longitudinal resistance $R_{xx}$ as a function of filling $\nu$ measured at $T=0.3$ K and $B = 0$.~\textbf{c.}~Hall resistance $R_{xy}$ for three perpendicular magnetic fields $B = 0.05, 0.4$ and 1.2 T for density being swept back and forth.~The red and blue arrows indicate directions of density sweep.~\textbf{d.}~$R_{xy}$ at three different fillings for $B$ swept back and forth, as indicated by the arrows.~A reversal of hysteresis is seen at $\nu=1.86$. The coercive fields are indicated as $B^+$ and $B^-$.~\textbf{e.}~Colorplot of $\Delta R_{xy}$, defined as the difference between the values of $R_{xy}$ for the opposite field sweeps, as a function of $\nu$ and $B$.~The change in colour shows the reversal of magnetization that accompanies the occupation of electrons in different $K$ and $K'$ valleys represented with the inset schematics.~In the bottom panel, the ratio of the magnitude of negative and positive coercive fields is plotted as a function of $\nu$.~Coercive fields are asymmetric in positive and negative $B$ and the asymmetry flips exactly when the magnetization is reversed.
}
\end{figure*}

\begin{figure*}[bth]
\includegraphics[width=1.0\textwidth]{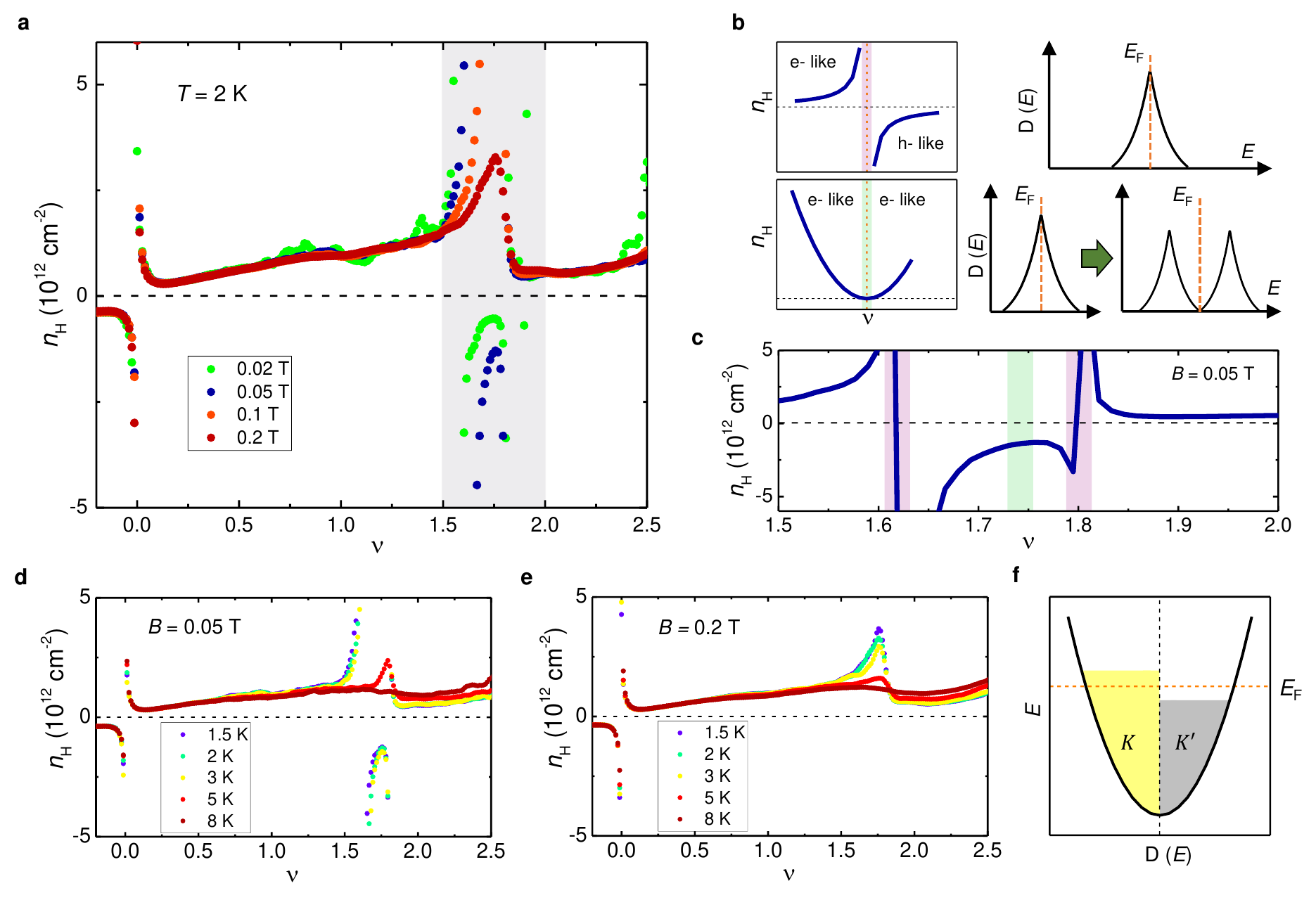}
\captionsetup{justification=raggedright,singlelinecheck=false}
\justify{\textbf{Fig.~2.~Fermi surface reconstructions and malleability of bands at $\boldsymbol{\nu<}~2$}.~\textbf{a.}~Hall density plotted as a function of $\nu$ for $B = 0.02, 0.05, 0.1$ and 0.2~T at $T = 2$ K.~Lifshitz transitions and reset of charge carriers are seen for $B = 0.02$ and 0.05 T , whereas only reset of charge carriers occur for $B = 0.1$ and 0.2~T.~\textbf{b.}~Expected behaviour of $n_H$ as a function of $\nu$ and the corresponding density of states D($E$) vs $E$ profile.
The purple line indicates a Lifshitz transition with a sign reversal in Hall density while the green line shows a reset where the Hall density reaches a minimum value without a sign change.~\textbf{c.}~$n_H$ vs $\nu$ at $B=0.05$ T for the shaded region in Fig.~2a.~Different colorbars are used to indicate Lifshitz transitions (purple) flanking the reset of carriers (green).~\textbf{d.-e.} Temperature dependence of Lifshitz transitions at $B = 0.05$ T and reset at $B = 0.2$ T.~\textbf{f.} Density of states showing a finite imbalance in occupation of states between $K$ and $K^\prime$ valleys leading to orbital magnetism.}
\end{figure*}

\begin{figure*}[bth]
\includegraphics[width=1.0\textwidth]{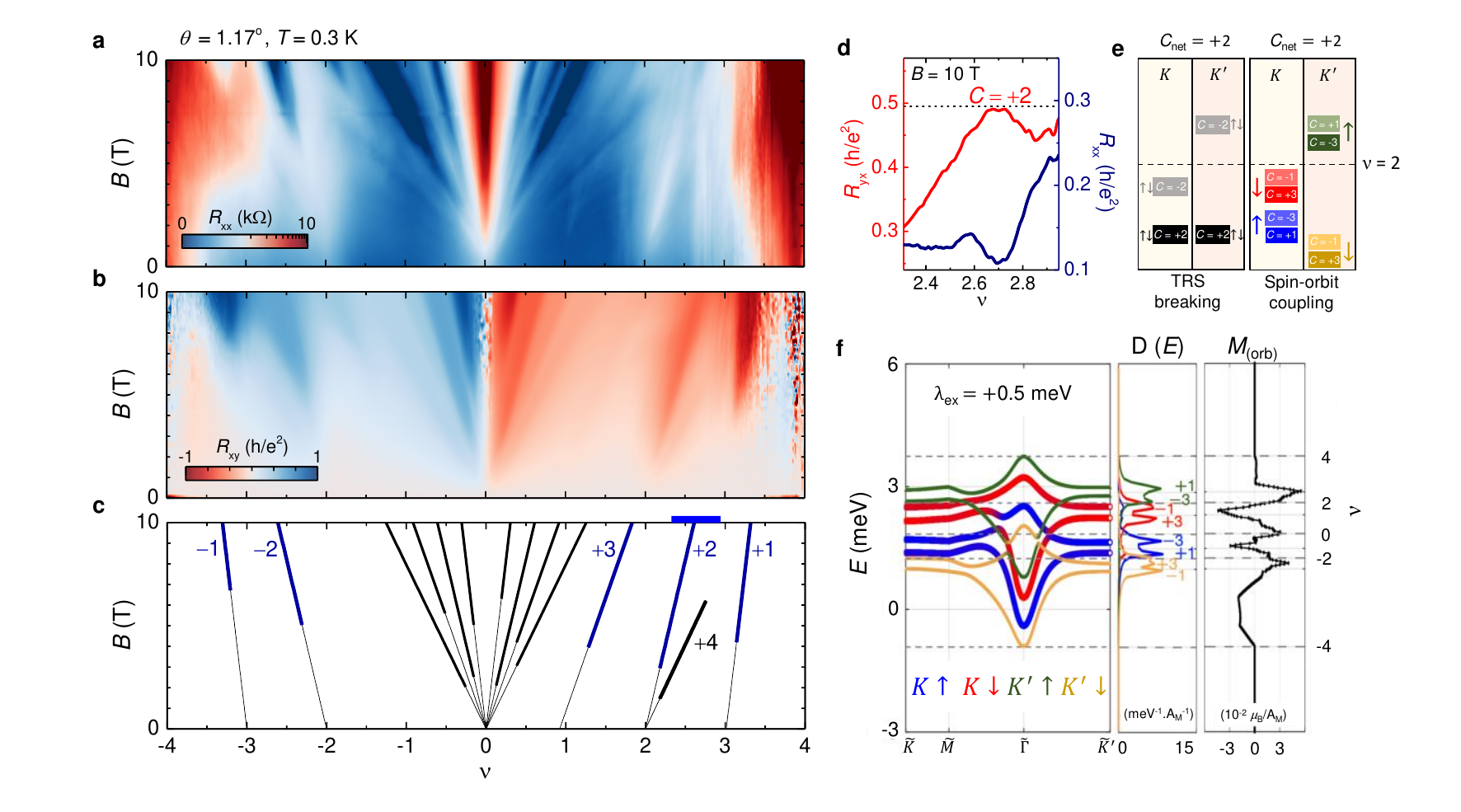}
\captionsetup{justification=raggedright,singlelinecheck=false}
\justify{\textbf{Fig.~3.~Quantized orbital Chern insulator at $\boldsymbol{\nu=}~2$ and possible Chern bands}.~\textbf{a-b.}~Landau fan diagram in longitudinal resistance $R_{xx}$ and transverse resistance $R_{xy}$ plotted as a function of $\nu$ for different $B$ up to 10 T.~\textbf{c.}~Fitting of the Diophantine equation along the minima in $R_{xx}$ and wedge like states in $R_{xy}$.~The slopes of the straight lines give the Chern numbers.~\textbf{d.}~$R_{xy}$ data at $B = 10$ T plotted for a small density range near $\nu = 2$ marked by the blue colorbar on the top axis in panel 3c.~$R_{xy}$ is quantized to $h/2e^2$ accompanied by a minima in $R_{xx}$ at $B = 10$ T.~\textbf{e.} Schematic of flat bands with different Chern numbers via valley polarizing $\mathcal{T}$-symmetry breaking and proximity-induced SOC.~The dark and light colors represent each spin-valley flavor's lower and upper bands, respectively.~In the presence of SOC, the spin-valley flavor degeneracy can be completely lifted, leading to 8 isolated bands for appropriate system parameters.~Blue(green) and red(yellow) indicate the up and down spin components for valley $K$($K^{\prime}$).~\textbf{f.} Schematic of valley Chern bands giving rise to $C_{\rm net} = +2$ at $\nu = +2$.~The band degeneracy can be lifted completely in the presence of SOC and a spin splitting exchange field that naturally accompanies a ferromagnetic spin polarized phase.~We have used the SOC parameters for graphene on WSe$_2$ following Ref.~\cite{PhysRevB.93.155104} together with the exchange field $\lambda_{\rm ex} = 0.5~\rm meV$ as summarized in the methods section.~The degeneracy split Chern bands lead to finite orbital magnetism $M_{\rm (orb)}$ that changes with the filling density $\nu$, shown here from $-4$ to $4$ in the right-most sub-panel.~We note the changing signs in the total orbital magnetization due to the relative filling of $K\downarrow$($K^{\prime}\uparrow$) bands near $\nu = +2$, suggesting that delicate changes in level ordering with carrier density due to Coulomb interactions can strongly impact the net magnetization. 
}
\end{figure*}


\begin{figure}[bth]
\includegraphics[width=0.8\columnwidth]{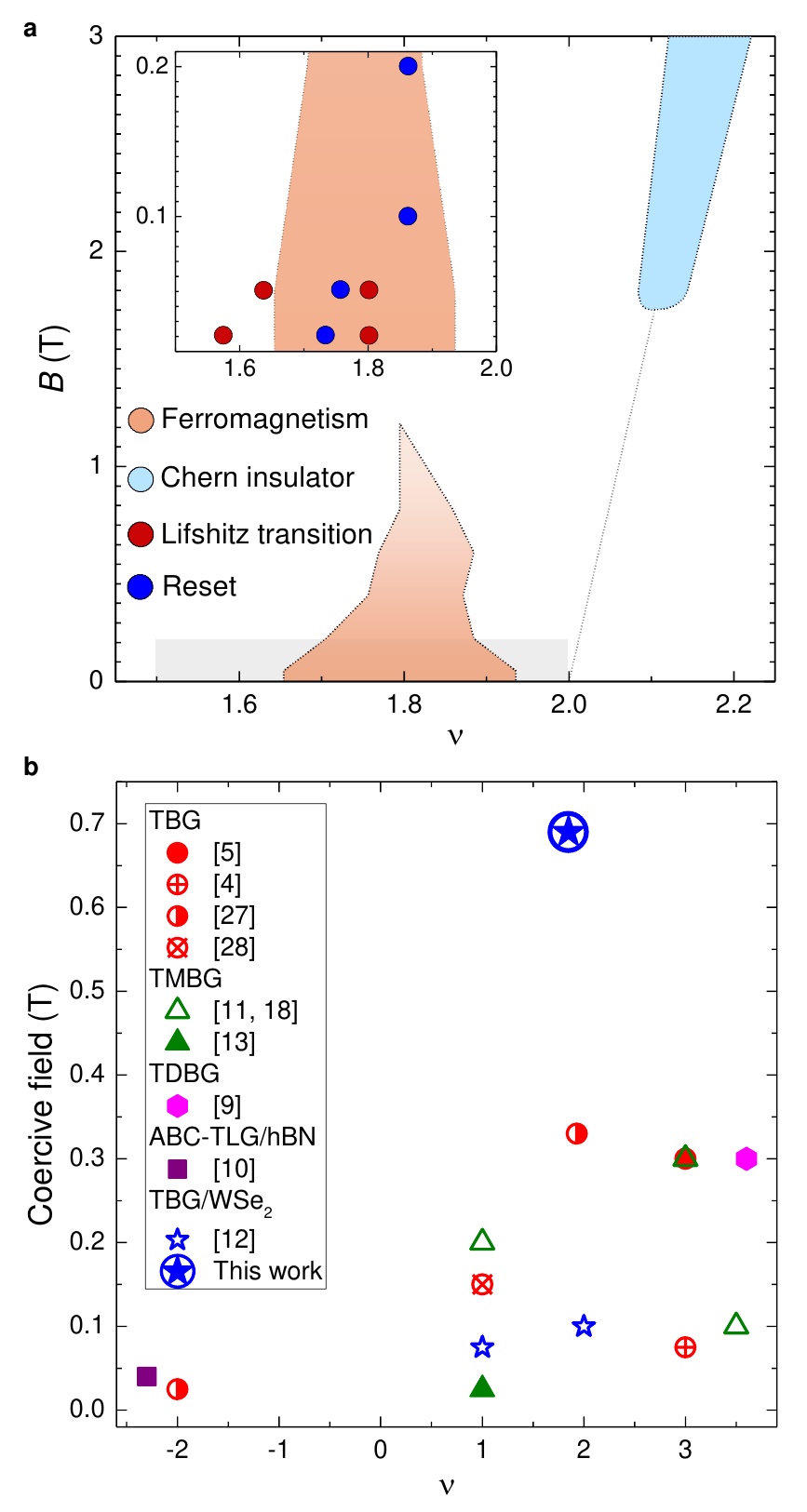}
\captionsetup{justification=raggedright,singlelinecheck=false}
\justify{
%
~\textbf{Fig.~4.~Summary of the various phases observed.}~\textbf{a.}~Four different colors are used to indicate the extent of the phases in the density and magnetic field subspace; Ferromagnetism (orange), Chern insulator (light blue), Lifshitz transitions (red) and reset (blue).~The inset shows low field phase diagram up to $B=0.2$~T for a density range of $\nu=1.5-2$ in the shaded region of the main panel.~The densities corresponding to Lifshitz transitions and reset are marked by red and blue circles respectively.~While ferromagnetism, Lifshitz transitions and reset occur below $\nu=2$, the Chern insulator emanates from exactly $\nu=2$ outside the ferromagnetic region.~\textbf{b.}~Coercive field as a function of $\nu$ for various reports on graphene-based moir\'{e} systems, clearly indicating the large coercive field observed in this work.~TBG, TMBG, TDBG, and TLG stand for twisted bilayer graphene, twisted monolayer bilayer graphene, twisted double bilayer graphene, and trilayer graphene, respectively.} 
\end{figure}

Topology of the Fermi surfaces and the density of states (DOS) at the Fermi level govern various competing orders in quantum materials~\cite{PhysRevB.101.125120, PhysRevB.102.125141}.~The formation of a broken-symmetry phase, such as a magnet, is often treated as an instability in the parent electron liquid phase, driven by singularities~\cite{PhysRevB.92.085423}.~For example, van Hove singularities (vHSs) which are associated with saddle points of energy dispersion in momentum space, feature strongly diverging DOS and favour localization of electronic states that stabilizes phases such as density waves, ferromagnetism and superconductivity~\cite{Polshyn2022, PhysRevB.102.245122, PhysRevB.101.125120, PhysRevB.102.125141, PhysRevB.92.085423}.~Contrary to these `local' vHSs, the whole electronic band in the magic-angle TBG is nearly flat, with a large, `global' DOS, that favor emergent correlated phases, including correlated insulators~\cite{cao2018correlated, cao2018unconventional, lu2019superconductors}, orbital magnets~\cite{lu2019superconductors, serlin2020intrinsic, sharpe2019emergent, PhysRevLett.123.096802, PhysRevLett.124.187601}, non-Fermi liquids~\cite{ghawri2022breakdown}, and Chern insulators~\cite{das2021symmetry, nuckolls2020strongly, wu2021chern, PhysRevB.85.195458, PhysRevB.102.041402}, typically at integer fillings of the moir\'{e} unit cell.~Experiments in TBG have shown that the inversion (C$_2$)~\cite{serlin2020intrinsic, sharpe2019emergent, PhysRevLett.124.166601} or time reversal ($\mathcal T$) symmetry breaking~\cite{das2021symmetry, wu2021chern} can lift the degeneracy of the flat bands and polarize spin-valley degrees of freedom leading to Chern insulators.~Reports of anomalous Hall effect (AHE) at zero magnetic field at moir\'{e} filling factors $\nu = 3$~\cite{serlin2020intrinsic, sharpe2019emergent} and most recently at $\nu =1, \pm2$~\cite{lin_spin-orbit, tseng2022anomalous, PhysRevLett.127.197701} necessitates a non-zero difference in the occupation of electronic states of the two valleys, that produces a finite Berry curvature.~While AHE and Chern insulators were most significantly observed in TBG samples aligned with a hexagonal boron nitride (hBN) layer~\cite{serlin2020intrinsic, sharpe2019emergent} or with the application of a large magnetic field~\cite{das2021symmetry, wu2021chern, nuckolls2020strongly}, spin-orbit coupling (SOC) can also drive topological order and symmetry-broken phases~\cite{lin_spin-orbit, choi2021correlation, bhowmik2022broken}.~Proximity-induced SOC can break C$_2\mathcal T$ symmetry at zero magnetic fields and polarize the charge carriers within a single valley~\cite{lin_spin-orbit, PhysRevB.102.235146}.~The interplay of SOC, interactions and topology, driven by the presence of vHSs has, however, remained largely unexplored~\cite{PhysRevB.96.235425}.~The presence of vHSs within the nearly flat moir\'{e} bands can lead to enhanced correlations 
that can be manifested in the emergence of new instabilities over a narrower range of tunable, non-integer moir\'{e} band fillings.\\

 In this work, we investigate the non-integer filling regime of the moir\'{e} band in TBG proximitized by tungsten diselenide (WSe$_2$).~We report signatures of valley polarization along with Fermi surface reconstructions that suggest a Stoner-like instability favoured by vHSs in the vicinity of $\nu<2$.~Significantly, the band reconstructions are malleable and can be tuned via a combination of carrier density and magnetic field.~Fig.~1a shows the schematic of our device consisting of a multilayer WSe$_2$ on magic-angle TBG encapsulated by two hBN layers.~The four-probe longitudinal resistance $R_{xx}$ as a function of filling $\nu$ at a  magnetic field $B = 0$ shows well-defined peaks at the charge neutrality point (CNP) $\nu = 0$ and half fillings $\nu = \pm2$ (Fig.~1b).~We estimate the twist angle to be $\theta\approx1.17^{\circ}$, consistent with the estimation from Landau fan diagram (Fig.~3).~The data presented in Fig.~1 were taken at a temperature $T=0.3$~K.~In Fig.~1c, the Hall resistance $R_{xy}$ at low $B$-fields, most strikingly, shows a hysteretic behavior over a wide range of fillings $\nu < 2$, as the Fermi energy is swept back and forth.~The hysteresis becomes narrower with increasing $B$-field and vanishes at $B\approx1.2$~T. ~Notably, while $R_{xy}$ shows zero-crossings at $B = 50$~mT, no sign change is observed for higher fields up to $B = 1.2$~T with $R_{xy}$ remaining negative for $B > 50$ mT.~This feature will be discussed in detail in Fig.~2.~It is evident that $R_{xy}$ strongly depends on the history of the sample in out-of-plane $B$-field training, and this leads to a non-zero $R_{xy}$ at $B = 0$ when the field is swept back and forth (Fig.~1d).~To our surprise, we find an abrupt sign change in the hysteresis of $R_{xy}$ at $\nu = 1.86$ (Fig.~1e).~The magnitude of the coercive field, where the hysteresis disappears, is about one order of magnitude higher than previous reports on moir\'{e} systems~\cite{serlin2020intrinsic, sharpe2019emergent, kuiri2022spontaneous, chen2022tunable, polshyn2020electrical, lin_spin-orbit}.~The large coercive field suggests a more robust ferromagnetic phase than in previous experiments and may also indicate domain wall pinning due to disorder and local inhomogeneities in the twist angle.~We expect that the coercive fields will couple more strongly with the orbital magnetic moments rather than the spin whose shifts in energy are typically of $\sim$~0.1~meV per Tesla.~The hysteresis in $R_{xy}$ with respect to both $\nu$ and $B$ suggests that the sample remains magnetized without any external magnetic influence.~We note that the measured $R_{xy}$ is much lower than the quantum of resistance ($h$/$e^2$, where $h$ is the Planck's constant and $e$ is electronic charge).~The width of the hysteresis in $B$-field changes as the carrier density is tuned in the vicinity of $\nu < 2$, as evident from the color plot in Fig.~1e, where we have plotted the difference in $R_{xy}$ for two opposite directions of field sweep, $\Delta R_{xy} = R_{xy}(\overleftarrow{B}) - R_{xy}(\overrightarrow{B})$, as a function of $B$ and $\nu$.~Surprisingly, we observe a significant asymmetry between positive ($B^+$) and negative coercive fields ($B^-$).~The sudden switching behavior of $R_{xy}$ as a function of density at $\nu = 1.86$ is accompanied by a reversal of the asymmetry between $B^+$ and $B^-$ coercive fields.~We quantify the latter feature using the parameter $\alpha$, defined as $\alpha = |B^-|/|B^+|$, obtaining $\alpha>1$ for $\nu\le1.85$ and $\alpha<1$ for $\nu>1.85$ (bottom panel of Fig.~1e).\\

~Ferromagnetism at $\nu=2$ is unexpected in TBG since a valley-polarized ground state is energetically unfavourable due to inter-valley Hund's coupling~\cite{PhysRevLett.124.166601}.~However, SOC in TBG leads to flat bands with non-zero Berry curvature within a single valley~\cite{lin_spin-orbit, PhysRevB.102.235146}.~Presence of proximity-induced SOC is confirmed by weak antilocalization measurements in our device (Supplementary Fig.~S7).~The reversal of magnetization is a strong evidence for spontaneous switching of valley polarization induced by tuning the carrier density.~A ferromagnet can be classified as spin or orbital, depending on whether the magnetization is due to spontaneous spin or valley polarization.~In an orbital Chern insulator, the magnetization jumps abruptly when the chemical potential crosses the Chern gap.~The edge state contribution is sufficient to change the sign of magnetization simply by tuning the density below the gap of an orbital Chern insulator~\cite{PhysRevLett.125.227702, polshyn2020electrical, PhysRevLett.49.405}. Therefore, the abrupt reversal of hysteresis indicates the dominance of orbital magnetism over the spin magnetism, and the energetically favourable ground state is solely determined by the gate voltages in weak magnetic fields.~The observation of a non-quantized $R_{xy}$, however, suggests that the ground states have partially valley polarized bands with unequal occupation of different valleys as a function of carrier density.~Partial valley polarization is not incompatible with the intervalley-coherent phases proposed in  literature~\cite{PhysRevX.10.031034, he2021competing, PhysRevB.102.035136} that can mitigate a fully valley polarized phase. We note, however, that the SOC terms by themselves do not mix electronic states from $K$ and $K^\prime$ and is not the microscopic origin for the intervalley-coherent phases.~The sign switching of valley polarizations as a function of density leads to an abrupt reversal of magnetization (Fig.~1e) indicating a clear phase transition point between these competing phases for magnetic fields below $\sim$~0.5~T.\\

Having established the orbital nature of the ferromagnet at $\nu < 2$, we now turn to the zero-crossings in $R_{xy}$ that accompany the hysteresis in $\nu$ at $B = 50$~mT (Fig.~1c).~The $\nu$-dependence of Hall density $n_H$ gives insights into the Fermiology of a system.~In Fig.~2a, $n_H = -(1/e)(B/R_{xy})$ is plotted as a function of $\nu$ at four different low $B$-fields, but at a higher temperature $T = 2$~K, where ferromagnetism disappears (Supplementary Fig.~S5) and the Hall data is independent of the direction of density sweep.~We observe a rich sequence of sign changes and resets in $n_H$ around $\nu<2$, particularly for the lowest fields 20~mT and 50~mT. Assuming a single particle energy band diagram for TBG, the DOS is expected to show a vHS around $\nu=2$ (see Fig. 2b, top right panel). As the Fermi energy is swept through the vHS, a Lifshitz transition is expected that changes the topology of the Fermi surface, flipping the sign of $n_H$ with a logarithmically diverging profile, as shown in the top left panel~\cite{kim2016charge}.~However, when the bands are malleable, as the Fermi energy approaches the peak in the DOS, it can reset the bands and produce a split DOS profile as shown in the bottom right panel~\cite{wu2021chern, bhowmik2022broken}.~This leads to a `reset' of charge carriers, where $n_H$ drops to a low value before rising again, without a sign change. Our experiments reveal a completely new set of phase transitions in comparsion to previous reports on magic-angle TBG, where a reset is typically observed near $\nu=2$~\cite{wu2021chern, saito2020independent,bhowmik2022broken}. A closer look at our data near $\nu<2$ at 50~mT shows two Lifshitz transitions that flank a reset, indicated by the colorbars in Fig. 2c.~We note that the vHS within the nearly flat bands can shift the density of states weights for small changes in the twist angle~\cite{leconte2019,Leconte_2022}.~Surprisingly, our experiments reveal an unprecedented tunability of the DOS with $B$-field and density, further validating the malleability of the TBG bands.~The Lifshitz transitions disappear at $B = 100$~mT, and $n_H$ shows a peak-like feature that decreases to zero and increases slowly (Fig.~2a).~Such a `reset' of charge carriers at a relatively higher field, with no additional Lifshitz transitions, indicates $B$-field-driven changes in the DOS of the  bands.~Fig.~2d-e show that these phase transitions become weaker and fade away with increasing temperature.~Remarkably, these distinct features in $n_H$ appear around the same density $\nu < 2$ where we have observed ferromagnetism at lower temperatures of $T\le1$~K.
~The flat band condition of on-site Coulomb interactions ($U$) dominating over the kinetic energy of the carriers, and the diverging DOS around $\nu< 2$ easily satisfy the Stoner criterion of ferromagnetism $UD(E_F) > 1$, where $D(E_F)$ is the DOS at the Fermi energy $E_F$~\cite{PhysRevB.92.085423, PhysRevLett.123.096802, PhysRevB.98.054515}.~We speculate that such a strong instability in the DOS at $\nu<2$ favours spontaneous valley polarization, leading to the observed AHE along with the switching of magnetization (Fig.~1d-e).~Our theoretical calculations discussed below show that spin polarization together with SOC assists valley polarization.~Thus, a valley polarized orbital magnet with a non-zero spin polarization should be favored over a valley polarized magnetic phase without a net spin polarization.~In Fig.~2f we illustrate the scenario where conventional Stoner spin polarizing ferromagnetic phase is accompanied by valley polarization where $K$ and $K^{\prime}$ valleys are unevenly occupied. 
\\

To gain further insights into the possible ground state at half-filling, we have measured $R_{xx}$ and $R_{xy}$ simultaneously in a $B$-field up to $B = 10$~T, at $T=0.3$~K.~A series of symmetry-broken Chern insulators in the form of minima in $R_{xx}$ and wedge-like features in $R_{xy}$ emerge from different fillings (Fig.~3a-b).~The Chern insulators can be characterized by fitting the Diophantine equation, $n/n_0 = C\phi/\phi_0 + s$, where $n_0$ is the density corresponding to one carrier per moir\'{e} unit cell, $C$ is the Chern number, $\phi$ is the magnetic flux per moir\'{e} unit cell, $\phi_0 = h/e$ is the flux-quantum, and $s$ is the band filling index or the number of carriers per unit cell at $B = 0$ T.
For sufficiently strong magnetic fields the four fold spin-valley degeneracy is completely lifted near the CNP: ($C, \nu$) = ($\pm1, 0$), ($\pm2, 0$), ($\pm3, 0$), ($\pm4, 0$).~In addition, we observe states emanating from different integers $\nu$ as ($C, \nu$) = ($+3, +1$), ($\pm2, \pm2$), ($+4, +2$), ($\pm1, \pm3$) (Fig.~3c).~The Chern insulator $C=2$ at $\nu=2$ is perfectly quantized to $h/2e^2$ at a high $B$-field (Fig.~3d).
~In TBG devices, such topological incompressible insulators have been described within the picture of isolated eight fold bands with broken $\mathcal T$ symmetry, where the Chern numbers of the bands are the same in the two valleys~\cite{das2021symmetry, nuckolls2020strongly, wu2021chern}, but opposite for valence and conduction bands (Fig.~3e).~The valley imbalanced filling of the bands results in the net Chern number observed, consistent with previous studies.~While a large $B$-field is usually used to break $\mathcal T$-symmetry, proximity-induced SOC in graphene due to WSe$_2$ breaks $C_2\mathcal T$ symmetry inherently at $B = 0$~\cite{PhysRevB.102.235146} and together with an exchange field it can generate spin-valley degeneracy-lifted bands (Fig.~3e).
Moreover, the spin-valley flavor resolved Chern numbers
can be tuned by varying the sublattice splitting energy and Ising SOC in TBG-WSe$_2$ systems. The values of exchange fields are expected to change with the degree of spin polarization at the onset of magnetism that depends on specific system parameters and Coulomb interactions, see supplementary information and Fig.~S9 that shows how different initial conditions for a mean-field self-consistent Hubbard model result in meta-stable spin polarized states at different carrier densities.~We illustrate in Fig.~3f, the spin-valley resolved band degeneracy lifting introduced by a finite exchange field of $\lambda_{\rm ex} = 0.5$~meV in the Hamiltonian that models the spin polarization of a ferromagnetic phase, together with a proximitized SOC term discussed in the methods section. %
The associated spin-valley resolved bands develop a well defined Chern number that will lead to a finite orbital moment when they are filled. We illustrate by using frozen bands how the total orbital moment evolves with filling density giving rise to a magnetization of 
the order of $\sim$10$^{-2}$~$\mu_{\rm B}/A_{\rm M}$ that can change its sign depending on the specific carrier density value. 
Since the orbital magnetization depends sensitively on the actual exchange field for the specific spin configuration, see Fig.~A1-A2,
it is expected that its integrated magnitude as well as the local values in experiments vary considerably with density, especially near the phase transition points. Experimentally, the orbital magnetization maps can reach local values as large as a few $\mu_{\rm B}/A_{\rm M}$~\cite{AFYorbital2021,Grover2022}.
In our experiments, an abrupt change in  orbital moment as a function of filling indicates that reordering of levels at $\nu=1.86$ takes place due to a close competition between the magnetic phases of opposite signs.\\

In Fig.~4a, we have presented a diagram with the summary of various phases discussed throughout this report.~The Lifshitz transitions and reset of carriers at $B=20$ and 50~mT occur at the densities near the extreme left boundary of the ferromagnetic domain as well as within the domain.~At $B\geq100$~mT the  Lifshitz transitions disappear and we find a second reset near the extreme right boundary of the ferromagnetic domain (inset in Fig.~4).~As evidenced from the diagram, ferromagnetism is accompanied by a series of Fermi surface reconstructions at the vHSs.~However, the observation of a Chern insulator emanating from an integer value of $\nu=2$ that lies outside the ferromagnetic domain in both density and magnetic field indicates that $\mathcal{T}$-symmetry breaking via high $B-$fields is distinct from the ferromagnetism observed at $\nu<2$.~Finally, in the context of a very high coercive field in our data, we have plotted coercive field reported in several moir\'{e} graphene systems at different $\nu$~(Fig.~4b).~The plot clearly indicates the coercive field observed in our work is the highest in comparison to other reports to date.\\

To summarize, our experiments have revealed a phase diagram of competing phases in TBG near its first magic angle which is indicative of vHSs within the quasi-flat bands, where proximity SOC plays a role in favoring the valley polarized solutions. This diverging DOS within the nearly flat bands of TBG reveals a finer internal structure of the bands that is manifested through multiple phase transitions as a function of temperature, magnetic fields and carrier densities in the vicinity of $\nu \sim 1.8$. 
Uncovering the underlying physics of the various quasi-degenerate, competing ground states will require an overarching theoretical analysis of strongly interacting many-body physics. Our primary findings of AHE and Fermi surface reconstructions are reported away from the usual commensurate filling of $\nu=2$.~The various features in our data including abrupt reversal of magnetization and non-quantized $R_{xy}$ are clear signs of orbital magnetism and partially valley polarized ground state, where both bulk and edge modes are expected to contribute to the transport.~The bulk transport may be affected by percolating conduction channels between topological domains of closely competing phases where external electric or magnetic fields can be used as control knobs to favor certain phases over the other.~Varying the twist angle between the graphene sheet and the WSe$_2$ can modify the proximity SOC strength that would in turn modify the phase diagram of the expected ground states. The high sensitivity of the electronic structure to experimental conditions makes it both a challenge to perform experiments reproducibly and at the same time provides an opportunity to explore the physics near multi-phase transition points where the electronic response functions will be particularly sensitive to external perturbations.
Whether the observed singularities within the flat bands fall under the category of higher-order vHSs, recently proposed in TBG and related systems~\cite{yuan2019magic, PhysRevResearch.4.L012013}, would need to be addressed in future works.~Another interesting direction would be to identify the connection, if any, between the valley-polarized Stoner magnet found in our work and the superconducting phase in the vicinity of $\nu = \pm2$~\cite{cao2018unconventional, lu2019superconductors, saito2020independent, stepanov2020untying, arora2020superconductivity}.

\section*{Acknowledgements}
We gratefully acknowledge the usage of the MNCF and
NNFC facilities at CeNSE, IISc. U.C. acknowledges funding from SERB via SPG/2020/000164 and WEA/2021/000005. 
Y.J.P. was supported by the Korean National Research Foundation grant NRF-2020R1A2C3009142 and D.L. was supported by grant NRF-2020R1A5A1016518, as well as
the Korean Ministry of Land, Infrastructure and Transport (MOLIT) from the Innovative Talent Education Program for Smart Cities.
J.J. was supported by the Samsung Science and Technology Foundation under project SSTF-BAA1802-06. We acknowledge computational support from KISTI through grant   
KSC-2021-CRE-0389 and the resources of Urban Big data and AI Institute (UBAI) at the University of Seoul and the network support from KREONET. 
K.W. and T.T. acknowledge support from JSPS KAKENHI (Grant Numbers 19H05790, 20H00354 and 21H05233).
\\

\section*{Author Contributions}
S.B. fabricated the device, performed the measurements and analysed the data. B.G. contributed to measurements and analysis of data.~Y.J.P., D.L. and J.J. performed the theoretical calculations.~S.D. and R.S. assisted in measurements.~K.W. and T.T. grew the hBN crystals.~A.G. advised on experiments.~U.C. supervised the project.~S.B., J.J. and U.C. wrote the manuscript, with inputs from all authors. 

\section*{Competing interests}
The authors declare no competing interests.

\bibliographystyle{naturemag}
\bibliography{Article}

\section*{METHODS}

\subsection*{Device fabrication}

The well known `tear and stack' method was used to assemble the heterostructure in this work.~Polypropylene carbonate (PPC) film coated on a polydimethylsiloxane (PDMS) stamp was used for picking up individual layers of hBN, WSe$_2$ and graphene.~The final device was etched into a multi-terminal  Hall bar by reactive ion etching using CHF$_3$/O$_2$ followed by electron-beam lithography, and thermal evaporation of Ohmic edge contacts and top gate using Cr/Au (5 nm/60 nm). WSe$_2$ layer with a thickness of $\sim$ 3 nm was exfoliated from bulk crystals procured from 2D Semiconductors.\\

\subsection*{Transport Measurements}
Electrical transport measurements were performed in a He$^3$ cryostat with a 10~T magnetic field and a cryogen-free, pumped He$^4$ cryostat with a 9~T magnetic field.~Magnetotransport measurements were carried out with a bias current of $10$~nA, using an SR830 low-frequency lock-in amplifier at $17.81$~Hz.~The carrier density in the system was tuned by the top gate.~The twist angle was estimated using the relation, $n_s=8\theta^2/\sqrt{3}a^2$ where $a = 0.246$ nm is the lattice constant of graphene and $n_s$ ($\nu = 4$) is the charge carrier density corresponding to a fully filled superlattice unit cell.~For the measurements of hysteresis in $R_{xy}$, Onsager reciprocity theorem~\cite{PhysRev.37.405, serlin2020intrinsic} was used, details of which are given in the supplementary information.

\subsection*{Proximity SOC in G/WSe$_2$}
The interlayer coupling, in particular the proximity SOC induced in the graphene sheet on top of a WSe$_2$ can be modeled by combining sublattice dependent site potential differences together with Rashba and intrinsic SOC terms~\cite{Koshino_tmd}.
In the following we briefly outline how the bands of graphene can be altered under the proximity SOC effects of WSe$_2$.
\begin{figure}[tbhp]
\begin{center}
\includegraphics[width=1.0\columnwidth]{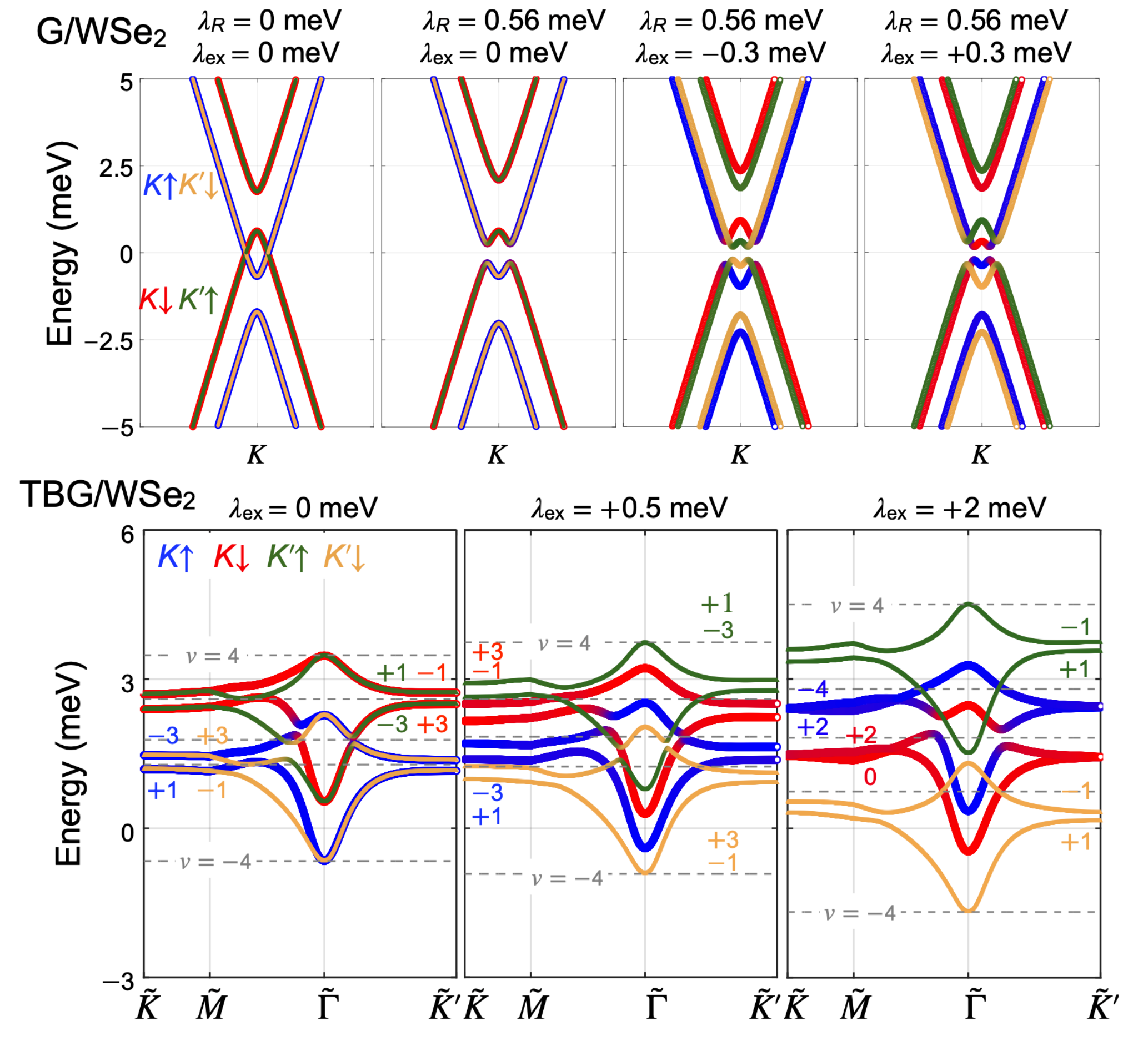}
\captionsetup{
justification=raggedright,singlelinecheck=false}    
\justify{
\textbf{Fig.~A1}
Exchange-field induced valley split bands of graphene on WSe$_2$ and TBG on WSe$_2$ calculated within the continuum model including SOC. For TBG we assume the magic twist angle $\theta = 1.05^{\circ}$ with a Fermi velocity of $\upsilon_F=1\times 10^6~\rm m/s$. 
The SOC parameters 
$\lambda_{I}^{A}$, $\lambda_{I}^{B}$, and $\lambda_{R}$,
$\lambda_{PIA}^{A}$, $\lambda_{PIA}^{B}$
use the default values proposed in Ref.~\cite{PhysRevB.93.155104} 
except when they are defined explicitly. 
{ The proximity SOC induced by the WSe$_2$ layer lifts the degeneracy of the spin up and down bands within each valley but the intervalley degeneracy between opposite spins of $K\uparrow$ ($K\downarrow$) and $K^{\prime}\downarrow$ ($K^{\prime}\uparrow$) bands is maintained.} An exchange field $\lambda_{\rm ex}$ that reflects the magnitude and sense of the spin polarization breaks the remaining degeneracy and can easily alter the ordering of the spin-valley flavours.
}
\label{fig:cont_bands}
\end{center}
\end{figure}

\begin{figure}[tbhp]
\begin{center}
\includegraphics[width=0.9\columnwidth]{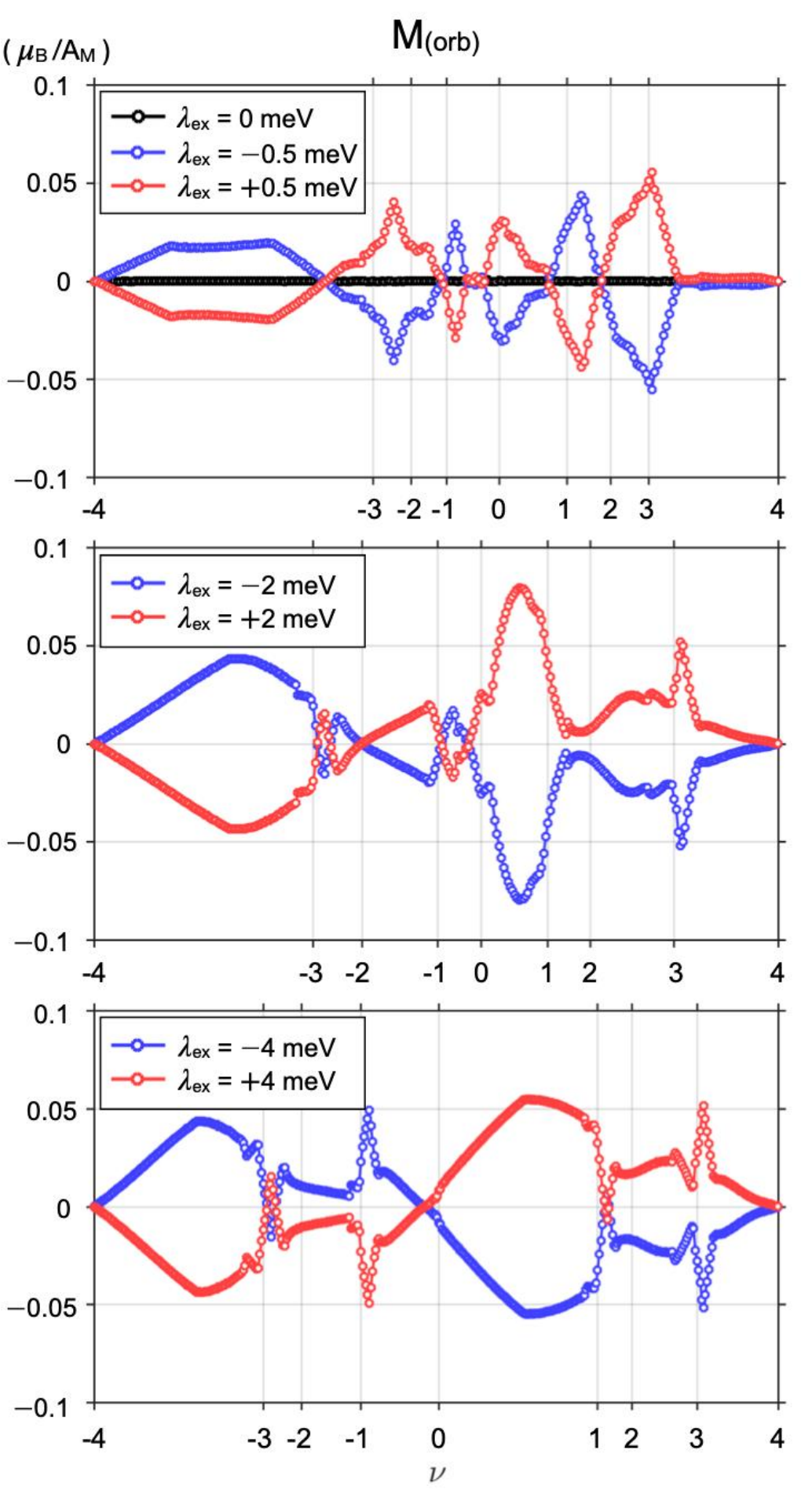}
\captionsetup{justification=raggedright,singlelinecheck=false}
\justify{
\textbf{Fig.~A2}
Orbital magnetization of the TBG-WSe$_2$ bands in Fig.~A1 using continuum model.
In the absence of the exchange field ($\lambda_{\rm ex}=0$), the orbital magnetization from the $K$ and $K^{\prime}$ valleys cancel each other leading to the zero total orbital magnetization (black). 
For the finite $\lambda_{\rm ex}= \pm 0.5,\ 2,\ 4$~meV, the orbital magnetization of each valley remains finite where the sign flip of the slope along with the band filling $v$ indicates the topological phase transition from positive to negative Chern number.
}
\label{fig:cont_OM}
\end{center}
\end{figure}
\subsection*{Continuum model bands}
We model the single-layer graphene Hamiltonian contacting a TMD layer through the staggered potential (U), exchange field (ex), intrinsic (I) and Rashba (R) SOC, and pseudospin inversion asymmetry (PIA).~\cite{PhysRevB.93.155104}
\begin{equation}
h_i \rightarrow h_i + h_U + h_I + h_{\rm ex} + h_R + h_{PIA}
\end{equation}
with
\begin{equation}
\begin{aligned}
h_U &= \Delta {\bf \sigma}_z \mathbf{1}_{\bf s}\mathbf{1}_{\bf \tau} \\
h_I &= \frac{1}{2} \left[\lambda_I^A(\sigma_z + \sigma_0) +\lambda_I^B(\sigma_z- \sigma_0)\right]\eta {\bf s}_z  \\
h_{\rm ex} &= \lambda_{\rm ex}  \mathbf{1}_{\bf \sigma} {\bf s}_z \mathbf{1}_{\bf \tau}\\
h_R &= \frac{\lambda_R}{2} (\eta {\bf \sigma}_x {\bf s}_y - {\bf \sigma}_y {\bf s}_x) \mathbf{1}_{\tau}\\
h_{PIA} &= \frac{a}{2} \left[\lambda_{PIA}^A(\sigma_z + \sigma_0) +\lambda_{PIA}^B(\sigma_z- \sigma_0)\right]\,\left( k_x {\bf s}_y  - k_y {\bf s}_x \right)  \mathbf{1}_{\bf \tau}\\
\end{aligned}
\end{equation}
where ${\bf \sigma}$ and ${\bf s}$ are Pauli matrices that represent the A/B sublattice and $\uparrow/\downarrow$ spin, and $\mathbf{1}$ is a $2\times2$ identity matrix.
 The simultaneous presence of a spin-splitting exchange field and a SOC term is known to introduce valley polarization. For instance, an intrinsic SOC captured through the Kane-Mele model together with sublattice staggering potential introduces unequal gaps at the $K$ and $K^{\prime}$ valleys that leads to an anomalous Hall effect by populating a given spin-valley band when the system is carrier doped.
Similarly, a bilayer graphene system subject to a Rashba SOC and spin-splitting exchange field is known to give rise to an anomalous Hall effect for appropriate system parameters~\cite{PhysRevB.82.161414}.
Both examples illustrate different mechanisms for the onset of an anomalous Hall effect when 
SOC is accompanied by an exchange field that separates the spin up and down bands in a ferromagnetic phase.
The Rashba SOC mixes spins but does not mix valleys.
The band structure of G/WSe$_2$ in Fig.~A1 illustrates how the two-fold low energy nearly flat bands are split into eight different bands due to spin-valley degeneracy lifting using a model system, 
whereas explicit mean-field calculations for the Hubbard model are presented in the supplementary information to illustrate the sensitivity of the calculated results to different initial conditions and carrier densities.
The high sensitivity of the orbital level ordering to spin configuration also leads to sensitive changes in the orbital moments with exchange field parameters. We illustrate in Fig. A2, the small changes in $\lambda_{\rm ex}$ up to $\sim$4~meV in magnitude which is sufficient to change the number of nodes crossing zero and therefore resulting in sign flips of the orbital magnetization slopes as a function of filling, in turn related with the spin-valley Chern numbers. 

Below, we write the pristine TBG continuum model Hamiltonian for one spin-valley flavor as 8$\times$8 matrix to emphasize the three dominant interlayer tunneling~\cite{Rafi2011PNAS} 
\begin{equation}
H_0^{\eta}({\bf k}) 
=
\begin{pmatrix}
h_{1}({\bf k}) & t_{12}({\bf k}_0)& t_{12}({\bf k}_+)&t_{12}({\bf k}_-)\\
t_{12}^{\dagger}({\bf k}_0)&h_{2}({\bf k}_0)&{\bf 0}&{\bf 0}\\
t_{12}^{\dagger}({\bf k}_+)&{\bf 0}&h_{2}({\bf k}_+)&{\bf 0}\\
t_{12}^{\dagger}({\bf k}_-)&{\bf 0}&{\bf 0}&h_{2}({\bf k}_-)
\end{pmatrix} 
\end{equation}
with
\begin{equation}
\begin{aligned}
h_{1}({\bf k}) &=  
\hbar\upsilon_F
\begin{pmatrix}
0  & (\eta k_x - i k_y) e^{i(+\frac{\theta}{2})}\\
(\eta k_x + i k_y) e^{i(-\frac{\theta}{2})} & 0
\end{pmatrix},\\
%
%
h_{2}({\bf k}) &=  
\hbar\upsilon_F
\begin{pmatrix}
0  & (\eta k_x - i k_y) e^{i(-\frac{\theta}{2})}\\
(\eta k_x + i k_y) e^{i(+\frac{\theta}{2})} & 0
\end{pmatrix}, \\
%
%
t_{12}({\bf k}_j) &=
\begin{pmatrix}
\omega'  & \omega e^{i(2\pi/3)j}\\
\omega e^{i(2\pi/3)j} & \omega'
\end{pmatrix}
\end{aligned}
\end{equation}
where ${\bf k}$ and ${\bf k}_j$ are the wave vectors measured from the $K^{(\eta)}$ of the layer 1 and 2 with the valley index $\eta=\pm1$, and ${\bf k}_j={\bf k}+{\bf G}_j$ with $j=0,+,-$ are connected with the three moire reciprocal lattice vectors, which adjusts the momentum difference between two different Dirac points from layer 1 and 2.
The $h_{1}$ and $h_{2}$ describe the Dirac bands of each layer where we use the Fermi velocity $\upsilon_F=1\times10^6$~m/s and the twist angle $\theta$. 
The $t_{12}$ term is the tunneling matrix between the two graphene layers with tunneling constants $\omega=0.12$~eV and $\omega'=0.0939$~eV. 
For the actual calculation, we use 392$\times$392 Hamiltonian matrices for each valley and each ${\bf k}$ point, which includes the hopping terms between the two spins, and use four sublattices, and 49 reciprocal lattice points.~\cite{PhysRevB.89.205414} The valley mixing terms are not included in our spin-orbit coupling models.

To understand the topological phase transition along with the band filling $\nu$, we obtain the orbital magnetization $M_{\rm (orb)}(\mu)$ as a function of chemical potential $\mu$~\cite{PhysRevLett.125.227702} using
\begin{equation}
\begin{aligned}
M_{\rm (orb)}(\mu) &= \sum_n  
\int \frac{d^2{\bf k}}{(2\pi)^2}\,
f(\mu-\varepsilon_n({\bf k}))\,
(\varepsilon_n + \varepsilon_{n^{\prime}}-2\mu)\\ 
&\ \times \left[\frac{e}{\hbar}{\rm Im}\sum_{n^{\prime}\neq n} \frac{\langle n| \partial_{k_x}H|n^{\prime}\rangle \langle n^{\prime}| \partial_{k_y}H|n\rangle }{(\varepsilon_n-\varepsilon_{n^{\prime}})^2}\right],
\end{aligned}
\end{equation}
where $f(E)$ is the Fermi-Dirac distribution, $\varepsilon_n$ and $|n\rangle$ are the eigen energy and vector of the $n$-th band.  
The Chern number $C = (2\pi\hbar/e) dM_{\rm (orb)}/d\mu$ can be estimated by the slope of the $M_{\rm (orb)}$ v.s. $\mu$ graphs. (See Fig. A2.)


\section*{Data availability}
The data that support the findings of this study are available from the corresponding author upon reasonable request.

\section*{Code availability}
The code that support the findings of this study are available from the corresponding author upon reasonable request.

\newpage
\clearpage

\onecolumngrid

\section*{SUPPLEMENTARY INFORMATION}

\section{Experimental Results}

{\begin{figure*}[bth]
\includegraphics[width=0.75\textwidth]{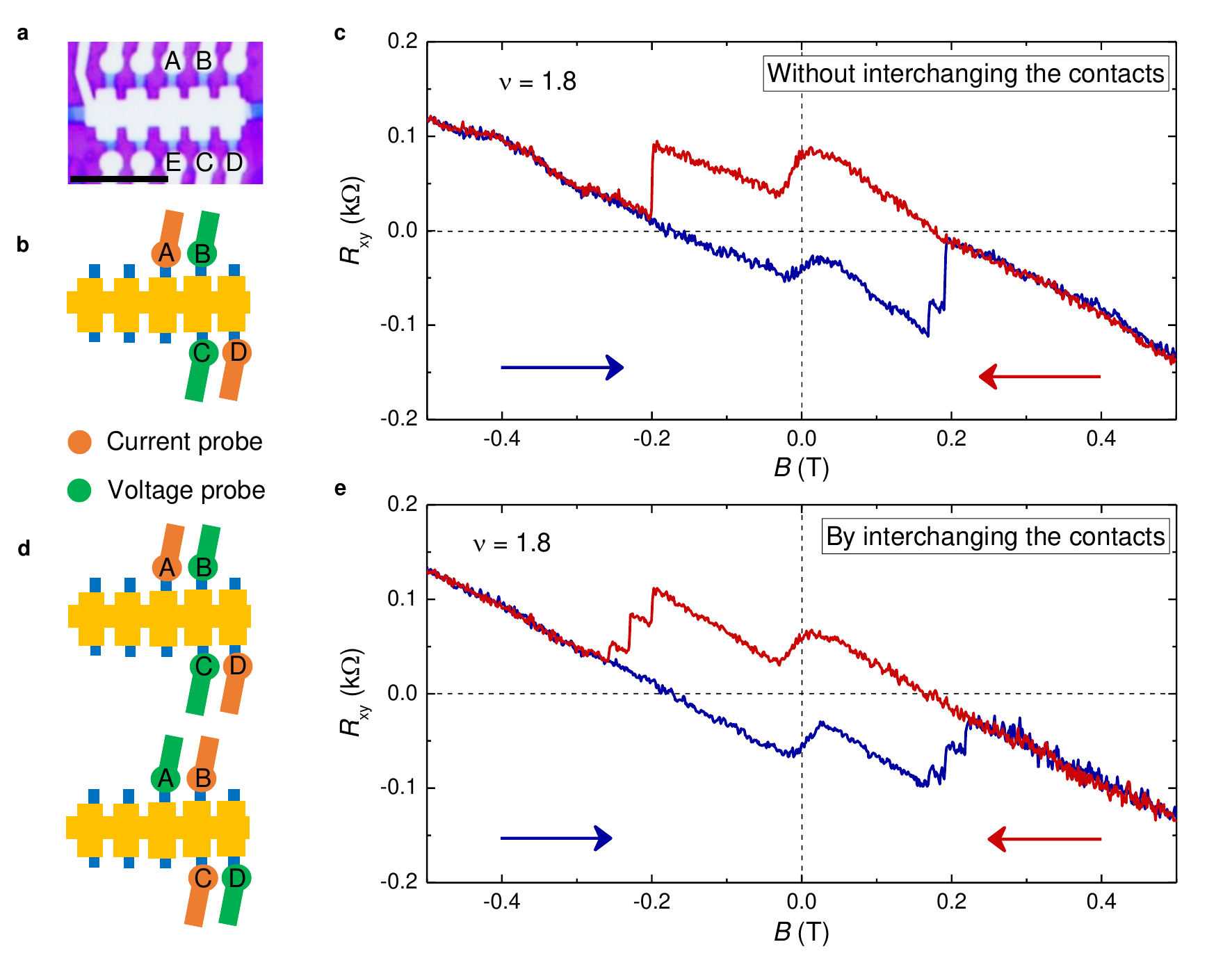}
\justify{\textbf{Supplementary Fig.}~S1.~\textbf{Hall resistance $R_{xy}$ measured as a function of magnetic field $B$ using Onsager reciprocity theorem~\cite{sample1987reverse, serlin2020intrinsic}}.~\textbf{a.}~Optical image of the device where A, B, C, D, and E are the contacts used throughout the measurements. The scale bar denotes 10~$\mu$m.~\textbf{b.-c.}~Measurement of $R_{xy}$ for $B$-field sweep at $\nu=1.8$ by driving current from contact A to D and measuring voltage between contact C and B.~Orange and green colours are used for current and voltage probes.~Blue and red colours in the data denote the direction of $B$-field sweep.~\textbf{d.-e.} First, $R_{xy}$ is measured as a function of $B$-field sweep by passing current from contact A to D and measuring voltage between contact C and B.~Second, $R_{xy}$ is measured as a function of $B$-field sweep by passing current from contact C to B and measuring voltage between contact A and D.~The antisymmetrized $R_{xy}$ is calculated using $R_{xy} = (R_{ADCB} - R_{CBAD})/2$.~The width and shape of the hysteresis obtained by measuring $R_{xy}$ using the above discussed two methods do not match perfectly.~The transverse voltage in an out-of-plane $B$ relies on the fact that $R_{xy}$ is antisymmetric in $B$ and it is accurately determined using $R_{xy} = \{R_{xy}(B)-R_{xy}(-B)\}/2$.~However, this method does not perfectly hold true for a ferromagnet where the dynamics of magnetic domains throughout the device may have different responses to the magnetic field.~For the precise determination of $R_{xy}$ below coercive field we have used Onsager reciprocity theorem and the ferromagnetism data presented in the paper are obtained using the reciprocity theorem, as detailed in (b).}
\end{figure*}

\begin{figure*}[bth]
\includegraphics[width=0.8\textwidth]{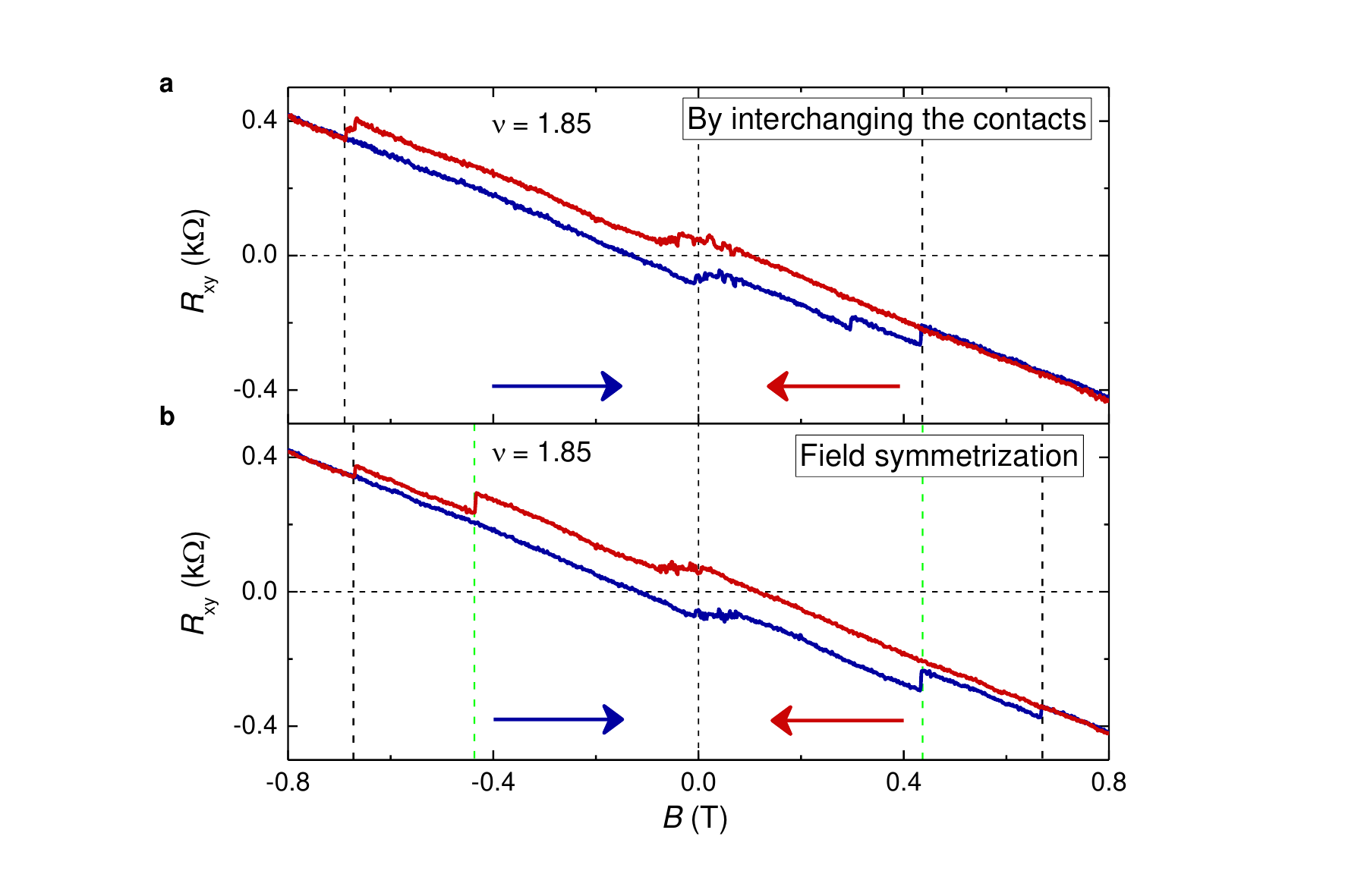}
\justify{\textbf{Supplementary Fig.}~S2. \textbf{Comparison of $R_{xy}$ using Onsager antisymmetrization and field antisymmetrization}.~\textbf{a.} $R_{xy}$ as a function of $B$ measured by interchanging current and voltage probes as discussed in Fig.~S1.~\textbf{b.} To compare $R_{xy}$ data resulting from different measurement configurations, we show the hysteresis at $\nu=1.85$ obtained from field antisymmetrized $R_{xy}(\overleftarrow{B}) = \{R_{xy}(\overleftarrow{B}) - R_{xy}(\overrightarrow{-B})\}/2$, $R_{xy}(\overrightarrow{B}) = \{R_{xy}(\overrightarrow{B}) - R_{xy}(\overleftarrow{-B})\}/2$ where $\overleftarrow{B}$ and $\overrightarrow{B}$ denote two opposite direction of $B$ sweep.~In this method, a significant reduction in the width of the hysteresis in $R_{xy}$ for $0.44\lesssim |B|\lesssim 0.67$ (shown by dashed green lines) indicates the asymmetry between positive and negative coercive fields.~However, the method discussed in \textbf{a.} better captures the dynamics of magnetic domains, which are asymmetric in magnetic field.~Therefore, we have used the Onsager anti-symmetrization method for the characterization of hysteresis in our device.}
\end{figure*}

\begin{figure*}
\includegraphics[width=0.8\textwidth]{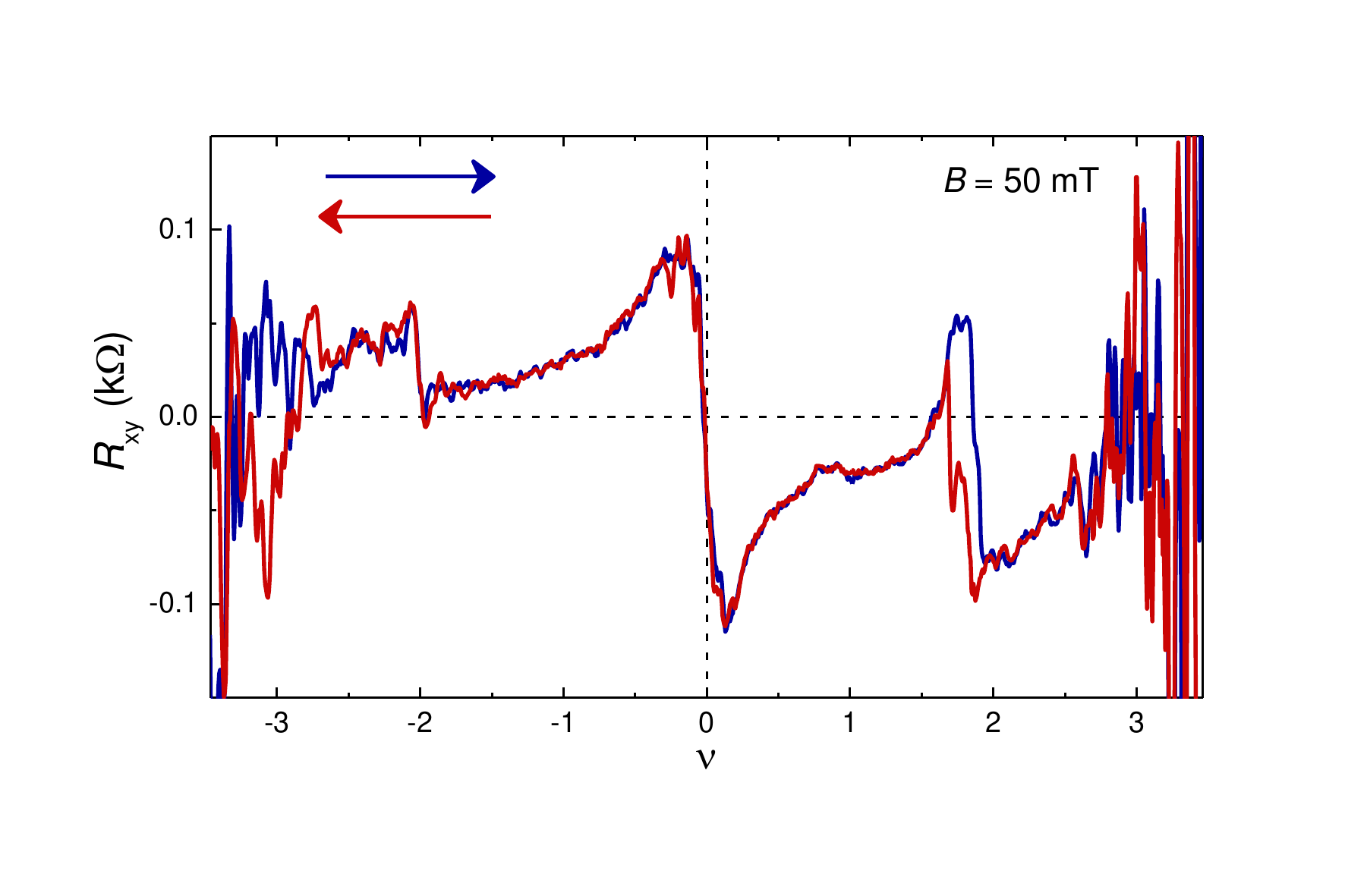}
\captionsetup{justification=raggedright,singlelinecheck=false}
\justify{\textbf{Supplementary Fig.}~S3. \textbf{Low-field Hall data for two opposite sweep directions of carrier density at $T=0.3$~K}.~$R_{xy}$ plotted as a function of $\nu$ at $B = 50$ mT shows a hysteresis loop with respect to density only in the vicinity of $\nu=2$.~Here, $R_{xy}$ is calculated using Onsager reciprocity theorem as discussed in Fig.~S1.}
\end{figure*}

\begin{figure*}
\includegraphics[width=0.8\textwidth]{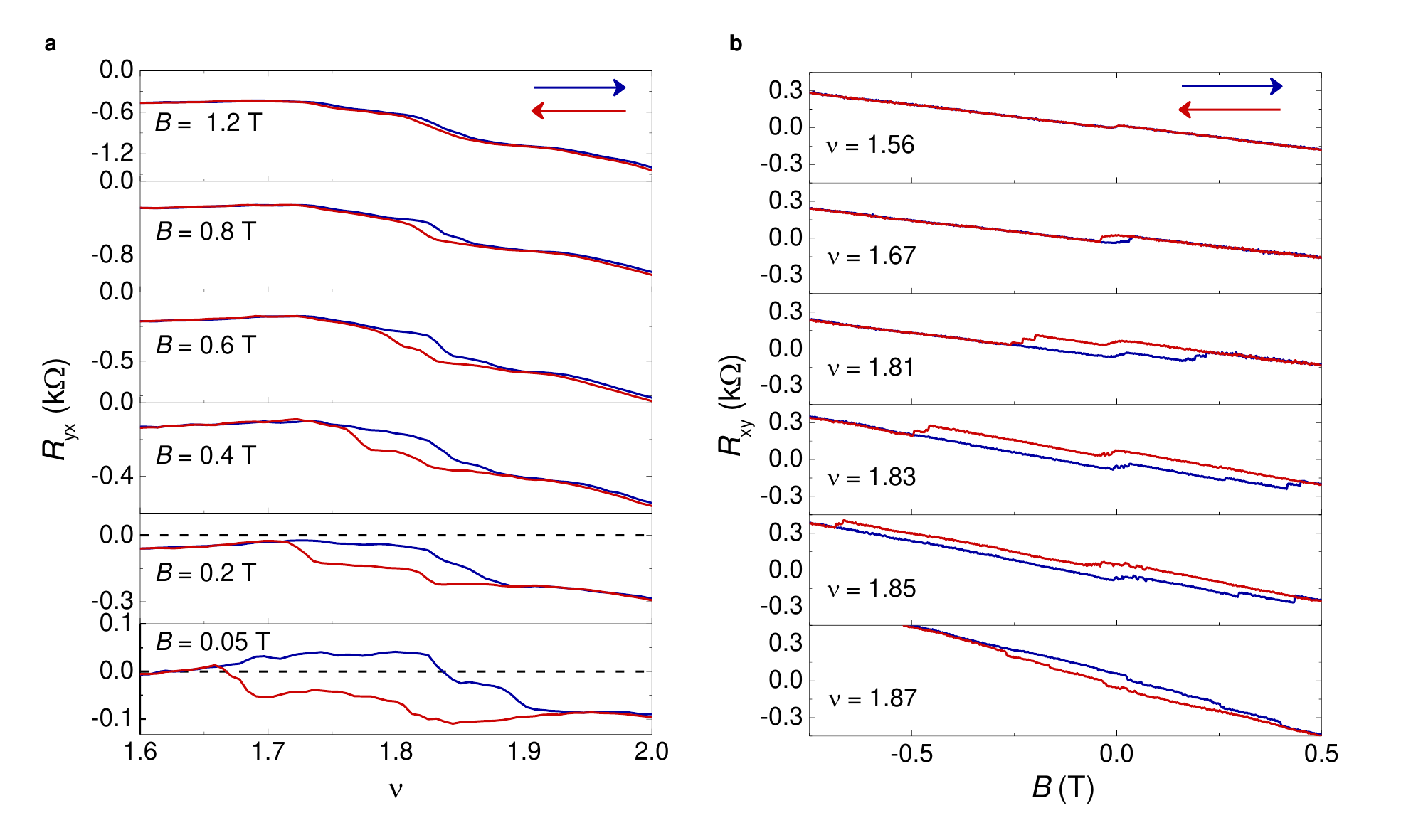}
\captionsetup{justification=raggedright,singlelinecheck=false}
\justify{\textbf{Supplementary Fig.} S4. \textbf{The evolution of hysteresis in density and magnetic field}.~\textbf{a.}~$R_{yx}$ shows hysteretic behaviour when the density is swept back and forth at low magnetic fields up to 1.2~T.~The hysteresis is strongly suppressed with increasing $B$.~$R_{yx}$ shows zero crossings only at $B = 0.05$~T and remains positive at other fields.~\textbf{a.}~The hysteresis in $R_{xy}$ with repect to out of plane $B$-field in plotted for different $\nu$ below $\nu = 2$.~The width of the hysteresis in $B$ is highly tunable with density and the reversal of the hysteresis in clearly seen.}
\end{figure*}

\begin{figure*}
\includegraphics[width=0.8\textwidth]{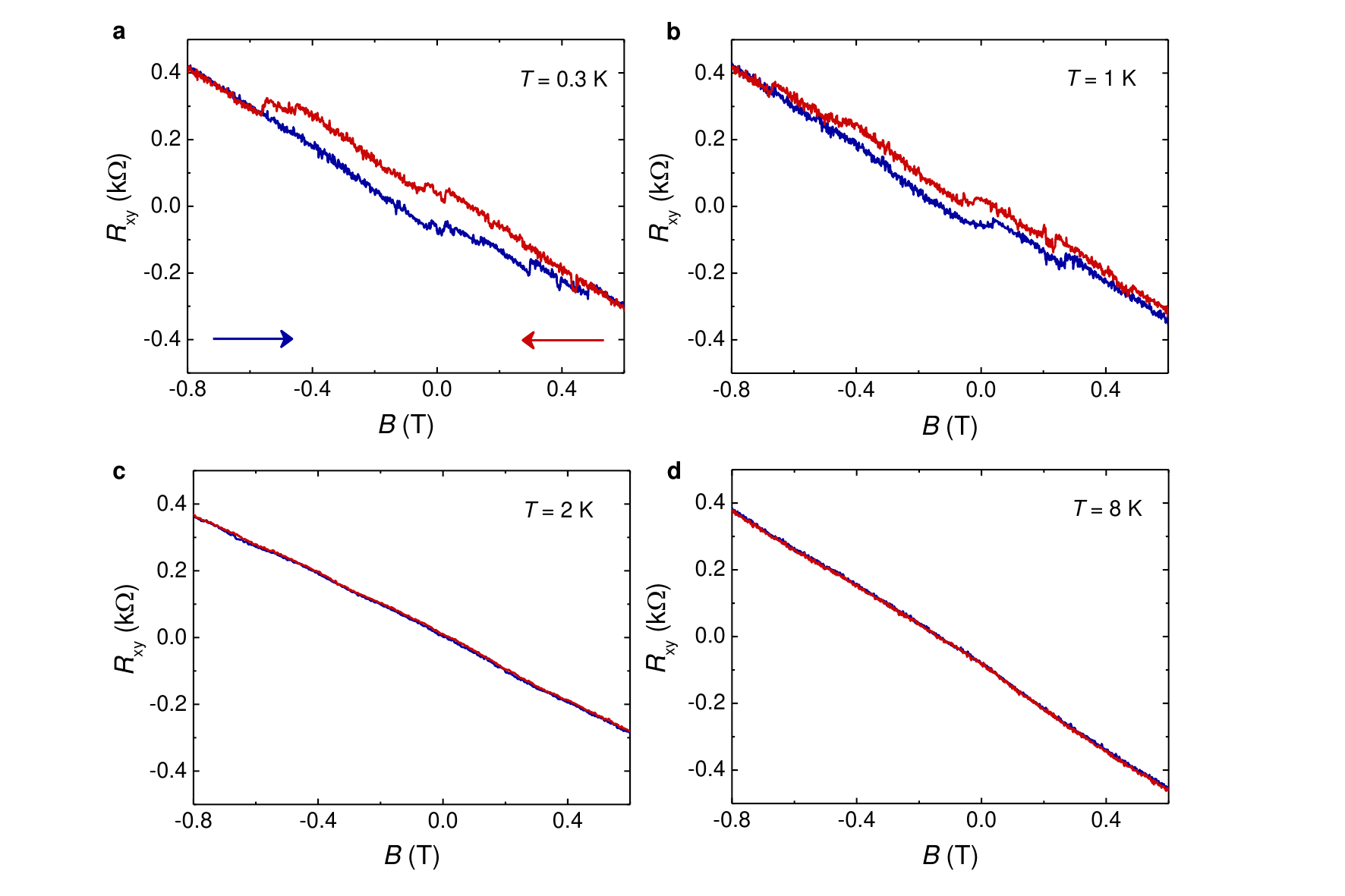}
\justify{\textbf{Supplementary Fig.}~S5. \textbf{Temperature dependence of the hysteresis} \textbf{a.-d.}~$R_{xy}$ as a function of $B$ for two opposite sweep directions is plotted for different temperatures.~The hysteresis becomes weaker with increasing temperature and disappears at $T = 2$~K.}
\end{figure*}
}

\begin{figure*}
\includegraphics[width=0.8\textwidth]{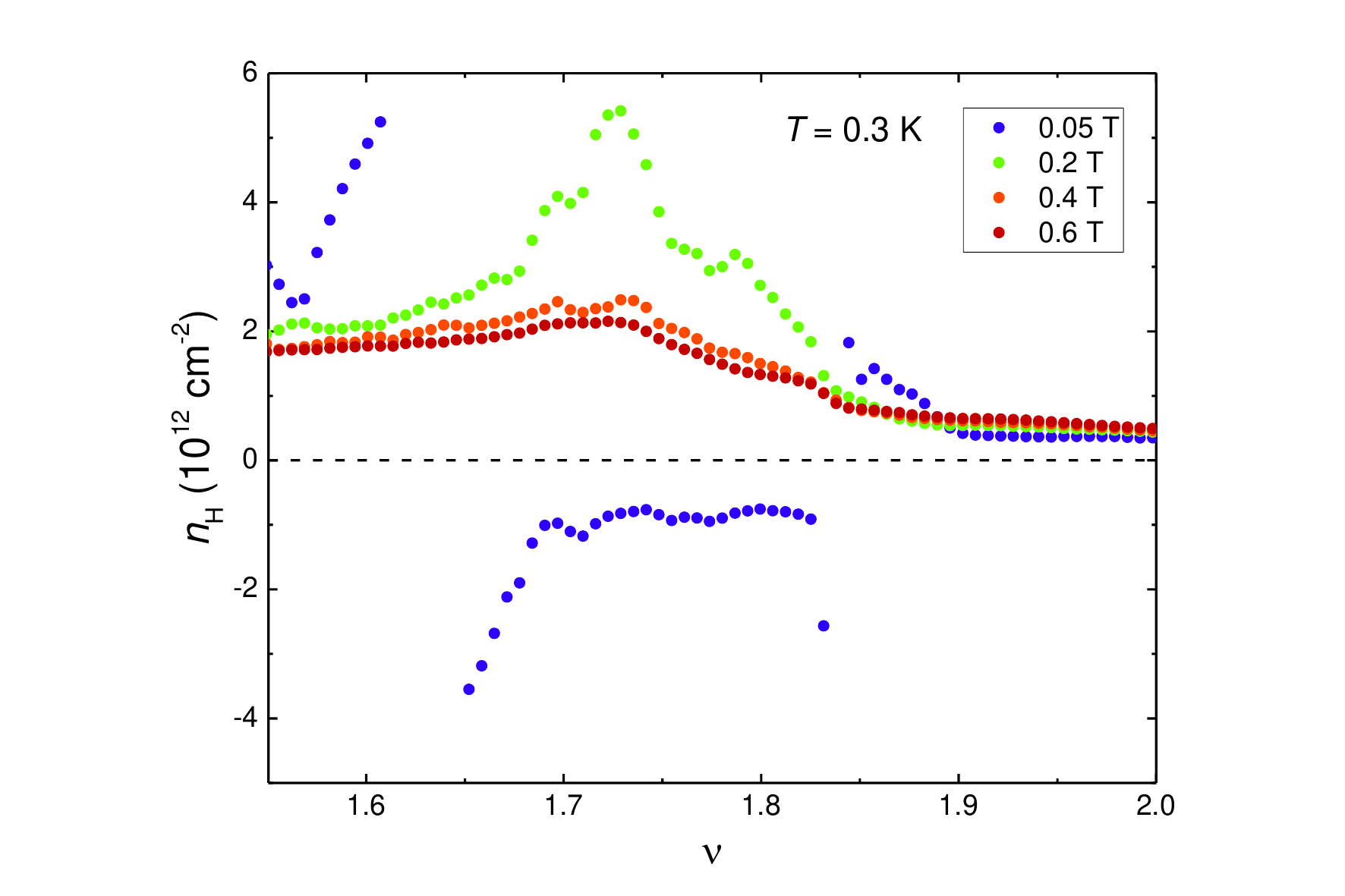}
\captionsetup{justification=raggedright,singlelinecheck=false}
\justify{\textbf{Supplementary Fig.}~S6. \textbf{Fermi surface reconstructions at $T = 0.3$~K}.~Hall density $n_H$ plotted as a function of $\nu$ for $B = 0.05, 0.2, 0.4$ and 0.6~T at $T = 0.3$ K.~Lifshitz transitions and reset of charge carriers are observed for $B = 0.05 T$ , whereas only reset of charge carriers occur for $B = 0.2-0.6$~T.}
\end{figure*}

\begin{figure*}
\includegraphics[width=0.8\textwidth]{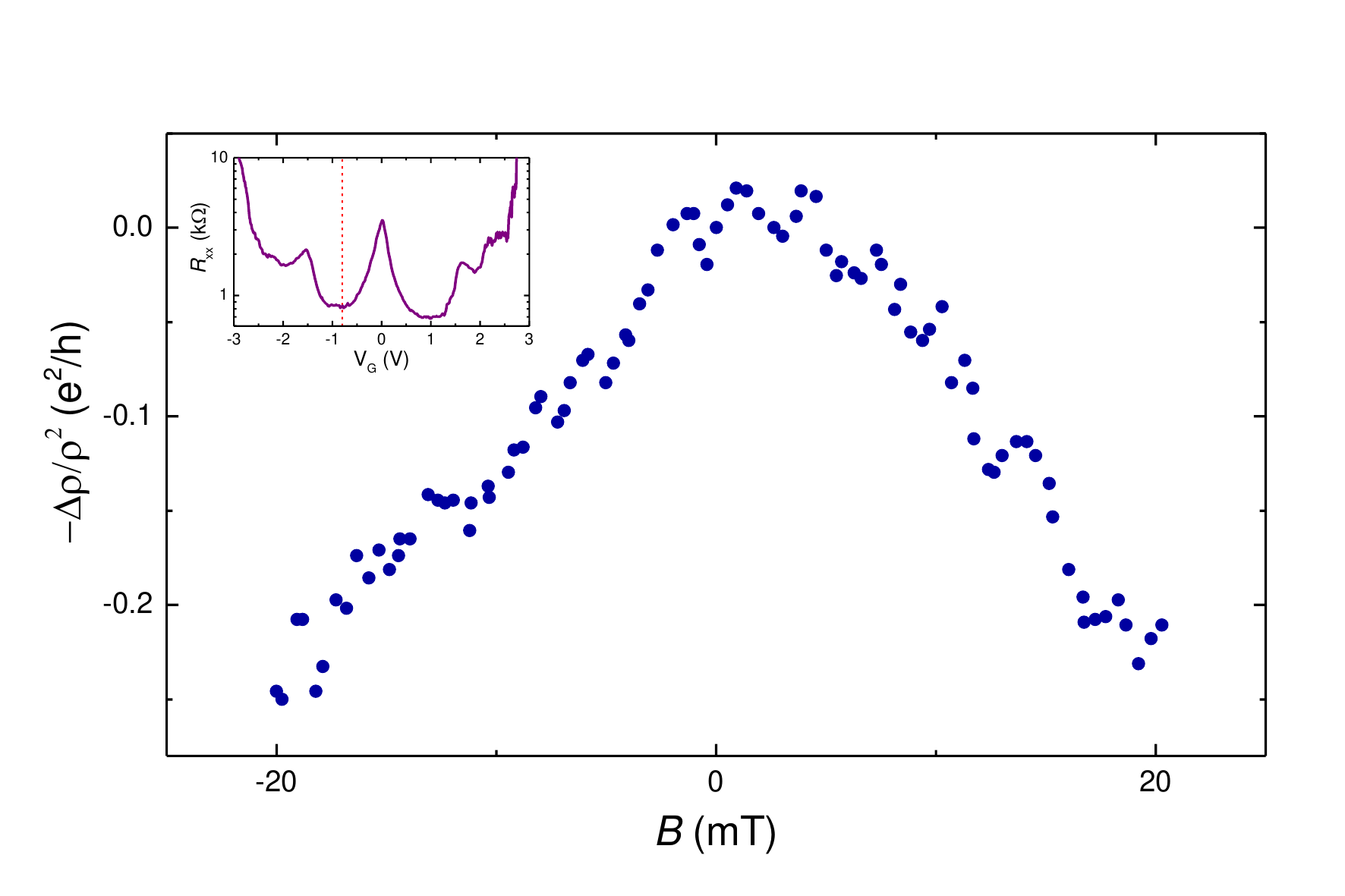}
\captionsetup{justification=raggedright,singlelinecheck=false}
\justify{\textbf{Supplementary Fig.}~S7. \textbf{Weak antilocalization data at $T=0.3$~K}.~The change in longitudinal magneto-conductivity ($-\Delta \rho/\rho^2$) with respect to $B = 0$ mT data, is plotted as a function of $B$ from -20~mT to 20~mT for a gate voltage $V_G = -0.8$~V~(shown by the dashed red line in the inset).~A peak in $-\Delta \rho/\rho^2$ at $B = 0$ mT indicates weak antilocalization and hence, a finite spin-orbit coupling in graphene layers proximity coupled to WSe$_2$.}
\end{figure*}

\begin{figure*}
\includegraphics[width=0.8\textwidth]{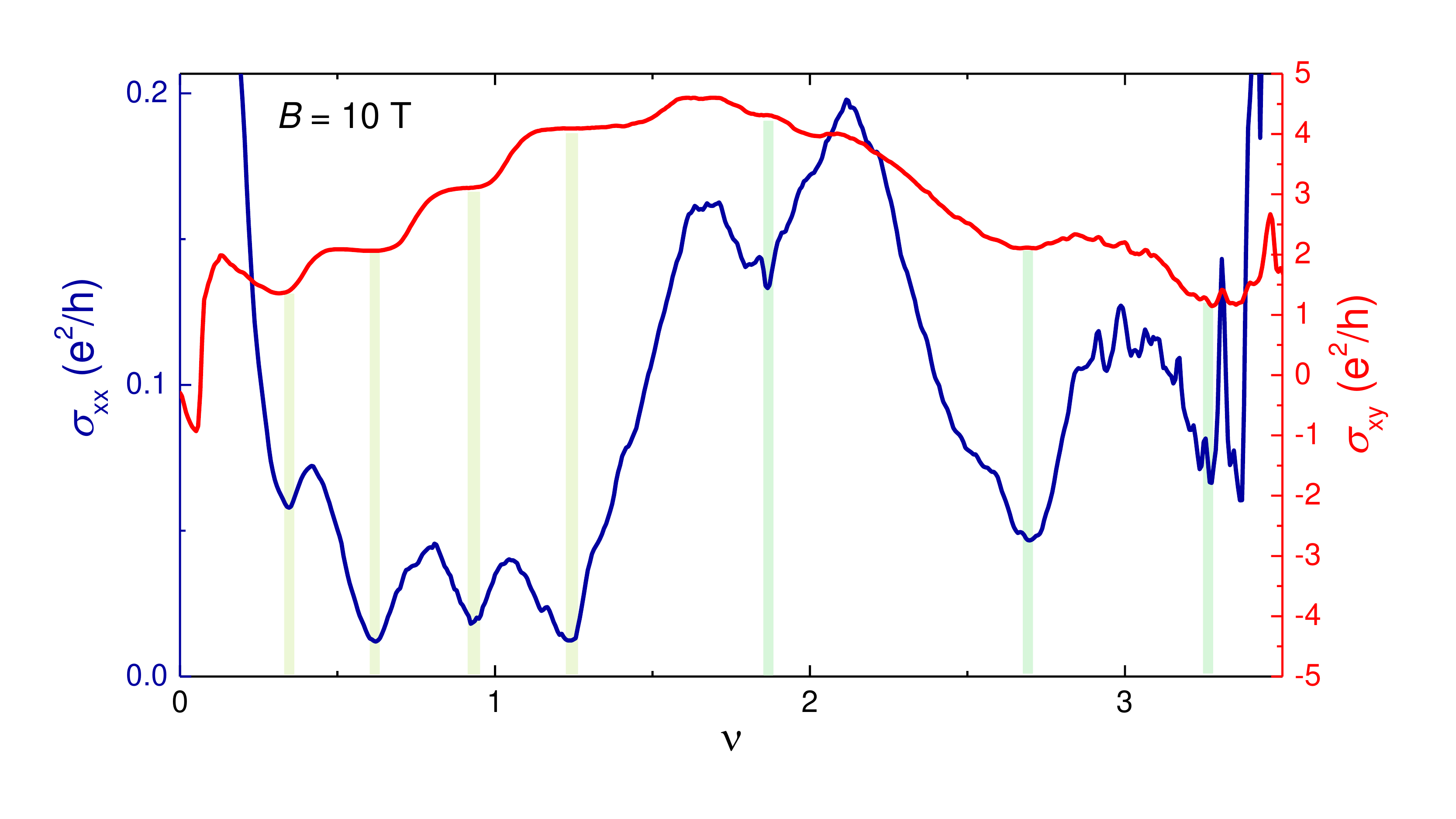}
\captionsetup{justification=raggedright,singlelinecheck=false}
\justify{\textbf{Supplementary Fig.}~S8. \textbf{Hall conductivity $\sigma_{xy}$ and longitudinal conductivity $\sigma_{xx}$ at $B = 10$ T}.~$\sigma_{xy}$ shows plateaus as $\sigma_{xy} = Ce^2/h$ associated with the minima in $\sigma_{xx}$. Different color bars (yellow for the CNP, green for $\nu = 1, 2, 3$) have been used to show the quantized states emanating from several partial fillings of flat bands.}\\
\end{figure*}

\newpage
\clearpage

\section{Spin-polarized phase in mean-field Hubbard model calculation}
\begin{figure}[tbhp]
\begin{center}
\includegraphics[width=0.7\columnwidth]{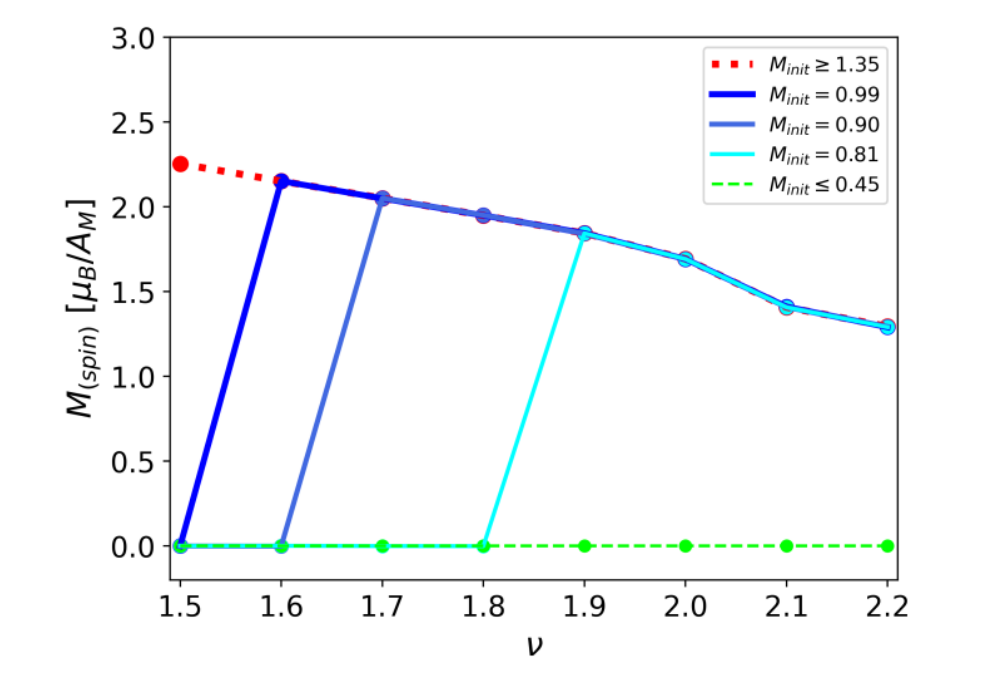}
\captionsetup{justification=raggedright,singlelinecheck=false}
\justify{
\textbf{Supplementary Fig.}~S9.
\textbf{Spin magnetization of converged states of Hubbard model calculation.}
The total spin magnetic moments of the moiré unit-cell was obtained from self-consistent spin density.
From $M_{init} \leq 0.45$ (Light green) and $M_{init} \geq 1.35$ (Red), the non-magnetic phase and the ferromagnetic phases are clearly identified. It is confirmed from the trend of magnetization change that the two phases become closer as the filling increases.
In $M_{init} = 0.81, 0.9, 0.99$ case, the states are separated between $\nu = 1.6$ and  $\nu = 1.9$ and converged to the two different phases manifested in lower and higher initial magnetization cases.
}
\label{fig:TB+U_mag}
\end{center}
\end{figure}
A mean field Hubbard model of the system introduces the spin-polarizes phase which further enhances the degeneracy lifting and therefore the valley polarization. The initial spin densities were assumed as uniformly distributed spin-ordered states of total magnetization in a unit moiré cell $M_{init} = 0 \sim4.5  [\mu_B/A_{Moire}]$. Interestingly, two stable convergence states of the non-magnetic and ferromagnetic phases were revealed according to the initial spin density. In addition, it was shown that the convergence phase transition at $\nu = 1.7$ under the initial magnetic condition of $M_{init} = 0.9$.

The tight-binding Hamiltonian of graphene on WSe$_2$ for the Hubbard model calculation inherits the proximity spin orbit coupling effect and sublattice dependent potentials due to the contacting layer. Following the formulations presented in Refs.~\cite{PhysRevB.93.155104} we consider the Hamiltonian

\begin{equation}
\begin{split}
    H =& H_0 
    + \sum_{i\in Bot,\sigma} \Delta \xi^i c^\dagger_{i,\sigma}c_{i,\sigma} \\
    & + \frac{2i}{3}\sum_{\langle i,j \rangle}^{Bot} \sum_{\sigma, \sigma'}\lambda_R c^\dagger_{i,\sigma}c_{j,\sigma'}(s^x_{\sigma,\sigma`}d^y_{i,j}-s^y_{\sigma,\sigma'}d^x_{i,j}) \\
    & + \frac{i}{3\sqrt{3}}\sum_{\langle \langle i,j\rangle \rangle}^{Bot} \sum_{\sigma, \sigma'}\lambda^i_I c^\dagger_{i,\sigma}c_{j,\sigma'}\nu_{i,j}s^z_{\sigma,\sigma'} \\
    & + \frac{2i}{3}\sum_{\langle \langle i,j\rangle \rangle}^{Bot} \sum_{\sigma, \sigma'}\lambda^i_{PIA} c^\dagger_{i,\sigma}c_{j,\sigma'}(s^x_{\sigma,\sigma`}d^y_{i,j}-s^y_{\sigma,\sigma'}d^x_{i,j}) \\
    & + \sum_{i,\sigma, \sigma' \neq \sigma} U \rho_{i,\sigma'} c^\dagger_{i,\sigma}c_{i,\sigma}
\end{split}
\end{equation}

Here, H$_0$ is non-interacting hopping model for magic-angle twisted bi-layer graphene with relaxation effects \cite{leconte2019},
and the second to fifth terms express spin-orbit interactions applied to the bottom layer graphene to effectively reflect the WSe$_2$ layers\cite{PhysRevB.93.155104}. The 1.7$^{\circ}$ commensurate tBG structure is used for this calculation with rescaling parameter $S'=1.608$ corresponding with the magic angle.
$\Delta$ is the induced orbital staggered potential, $\lambda_I$ is the intrinsic spin-orbit couplings, $\lambda_R$ is the Rashba spin-orbit coupling and $\lambda_{PIA}$ is the pseudospin-inversion-asymmetry(PIA) spin-orbit terms. We used $\Delta$ = 0.54 meV and $\lambda_R$ = 0.56 meV. In the case of $\lambda_I$ and $\lambda_{PIA}$, the values are different for each sublattice and are $\lambda^A_I$=$-$1.22~meV, $\lambda^B_I$=1.16~meV, $\lambda^A_{PIA}$ = $-$2.69~meV, $\lambda^B_{PIA}$ = 2.54~meV. The last term considers the mean-field Hubbard model. The spin density $\rho_{i,\sigma}$, considering the neutral density $n^0$ obtained from the Non-interacting Hamiltonian, is defined as $\rho_{i,\sigma} \equiv \langle c^\dagger_{i,\sigma}c_{i,\sigma} \rangle - n^0_{i,\sigma}$. The self-consistent calculations were performed on $5 \times 5$ Monkhorst-Pack grids for U = 5 eV.



\end{document}